\documentclass[usenatbib]{aa}

\usepackage{comment}
\usepackage{placeins}
\usepackage{graphicx}
\usepackage{amsmath,bm}
\usepackage{wasysym}
\usepackage{xspace}
\usepackage{xcolor}
\usepackage{soul}
\usepackage{natbib}
\usepackage{txfonts}
\usepackage{adjustbox}
\usepackage[version=4]{mhchem}
\bibpunct{(}{)}{;}{a}{}{,}
\nolinenumbers
\usepackage{txfonts,textcomp}

\usepackage[colorlinks=true,allcolors=blue]{hyperref}
\usepackage{etoolbox}

\makeatletter
  \patchcmd{\NAT@citex}
    {\@citea\NAT@hyper@{
      \NAT@nmfmt{\NAT@nm}
      \hyper@natlinkbreak{\NAT@aysep\NAT@spacechar}{\@citeb\@extra@b@citeb}
      \NAT@date}}
    {\@citea\NAT@nmfmt{\NAT@nm}
    \NAT@aysep\NAT@spacechar\NAT@hyper@{\NAT@date}}{}{}

  \patchcmd{\NAT@citex}
    {\@citea\NAT@hyper@{
      \NAT@nmfmt{\NAT@nm}
      \hyper@natlinkbreak{\NAT@spacechar\NAT@@open\if*#1*\else#1\NAT@spacechar\fi}
        {\@citeb\@extra@b@citeb}
      \NAT@date}}
    {\@citea\NAT@nmfmt{\NAT@nm}
    \NAT@spacechar\NAT@@open\if*#1*\else#1\NAT@spacechar\fi\NAT@hyper@{\NAT@date}}
    {}{}
\makeatother

\makeatletter
\AtBeginDocument{%
  \nolinenumbers                
  \ifdefined\makeLineNumber     
    \global\let\makeLineNumber\relax
  \fi
  \@ifundefined{linenumbers}{}{%
    \renewenvironment{linenumbers*}{}{}%
  }
  \@ifundefined{switchlinenumbers}{}{%
    \renewcommand{\switchlinenumbers}{}%
  }
}
\makeatother

\usepackage{soul}
\usepackage{adjustbox}

\newcommand{\Omegam}{\Omega_{\mathrm{m}}}
\newcommand{\Omegal}{\Omega_{\mathrm{\Lambda}}}
\newcommand{\LCDM}{\mathrm{\Lambda CDM}}

\newcommand{\kms}{\,{\rm km}\,{\rm s}^{-1}}

\newcommand{\Mpc}{\,{\rm Mpc}}

\bibpunct{(}{)}{;}{a}{}{,} 

\usepackage{txfonts,textcomp}
\usepackage[colorlinks=true,allcolors=blue]{hyperref}
\usepackage{etoolbox}

\begin{document}

   \title{SHAPE. I. A SOM-SED hybrid approach for efficient
   galaxy parameter estimation leveraging JWST}\authorrunning{Z. Wang et al.}

   \author{Zihao Wang \inst{1}
       \and
           Tao Wang \inst{1,2}\thanks{E-mail: \href{mailto:taowang@nju.edu.cn}{taowang@nju.edu.cn}}
       \and
           Ke Xu \inst{1,2} 
        \and
           Hanwen Sun \inst{1,2} 
        \and
           Ruining Tian \inst{1}
        \and
           Qi Hao \inst{1,2}
    }

   \institute{School of Astronomy and Space Science, Nanjing University, Nanjing, Jiangsu 210093, China
         \and
             Key Laboratory of Modern Astronomy and Astrophysics, Nanjing University, Ministry of Education, Nanjing 210093, China\\
             }
    \date{Received Month DD, Year; accepted Month DD, Year}

 
  \abstract
   {
   With the launch and application of next-generation ground- and space-based telescopes, astronomy has entered the era of big data, necessitating more efficient and robust data analysis methods.
   Most traditional parameter estimation methods are unable to reconcile differences between photometric systems. Ideally, we would like to optimally rely on high-quality observations data provided by, e.g., JWST, for calibrating and improving upcoming wide-field surveys such as the China Space Station Telescope (CSST) and \textit{Euclid}.
   To this end, we introduce a new approach (SHAPE, SOM-SED Hybrid Approach for efficient Parameter Estimation) that can bridge different photometric systems and efficiently estimate key galaxy parameters, such as stellar mass ($M_\star$) and star formation rate (SFR), leveraging data from a large and deep JWST/NIRCam and MIRI survey (PRIMER). As a test of the methodology, we focus on galaxies at z $\sim 1.5-2.5$. To mitigate discrepancies between input colors and the training set, we replace the default SOM weights with stacked SEDs from each cell, extending the applicability of our model to other photometric catalogs (e.g., COSMOS2020). By incorporating a SED library (SED Lib), we apply this JWST-calibrated model to the COSMOS2020 catalog. Despite the limited sample size and potential template-related uncertainties, SOM-derived parameters exhibit a good agreement with results from SED-fitting using extended photometry. 
   Under identical photometric constraints from CSST and \textit{Euclid} bands, our method outperforms traditional SED-fitting techniques in SFR estimation, exhibiting both a reduced bias (-0.01 vs. 0.18) and a smaller $\sigma_{\rm NMAD}$ (0.25 vs. 0.35).
   With its computational efficiency capable of processing $10^6$ sources per CPU per hour during the estimation phase, this JWST-calibrated estimator holds significant promise for next-generation wide-field surveys.
}

   \maketitle
%

\section{Introduction} \label{sec:intro}

\begin{figure*}
    \centering
    \includegraphics[width=0.9\linewidth]{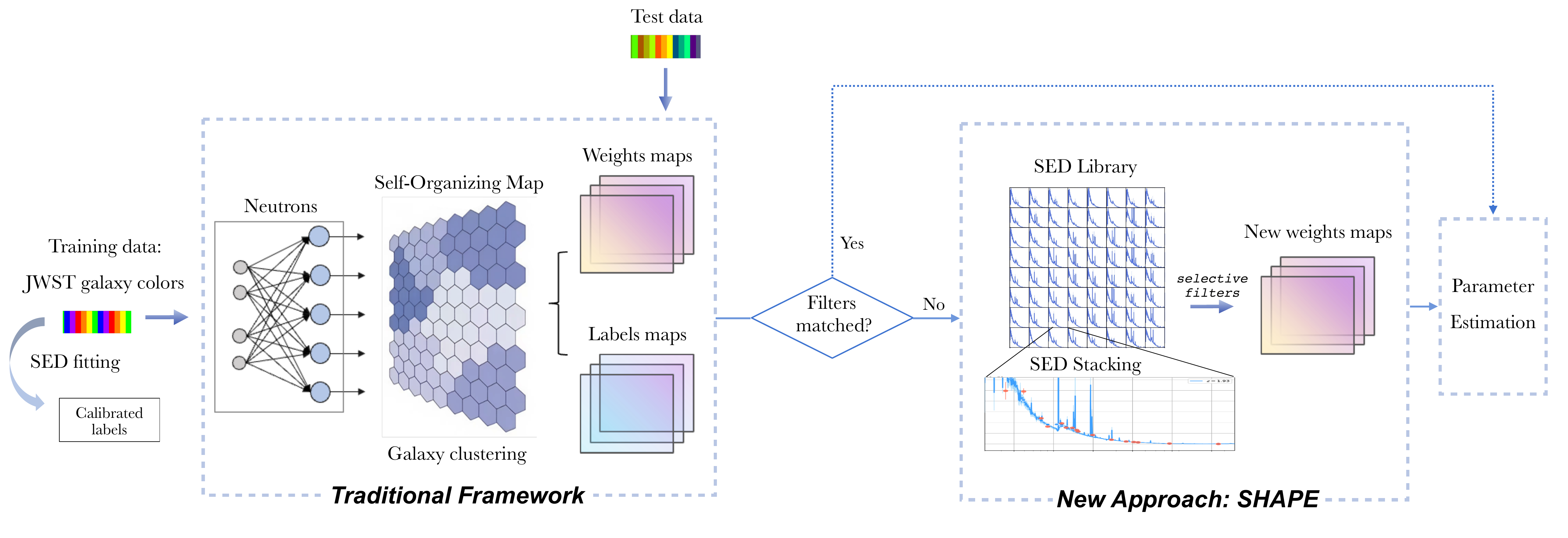}
    \caption{Schematic diagram of SHAPE model. This method employs a SOM to cluster galaxies in the training set and assigns each SOM cell a representative average SED. When the photometric filters of the test set match those of the training set, galaxies can be directly projected onto the SOM for parameter estimation. Otherwise, the galaxy is matched to the SED library (SED Lib) constructed from the SOM to determine its physical parameter.}
    \label{fig:process}
\end{figure*}

The fundamental physical parameters of galaxies, such as stellar mass (${M_\star}$) and star formation rate (SFR), are crucial for understanding galaxy formation and evolution. These parameters serve as key diagnostics for tracing the diverse evolutionary pathways of galaxies across cosmic time \citep{Madau_2014,Kennicutt_2012,Somerville_2015}. Consequently, the accurate and efficient determination of these properties has remained a long-standing challenge in extragalactic astronomy.

In the absence of spectroscopic data, traditional methods infer galaxy properties through spectral energy distribution (SED) fitting, which employs stellar population synthesis models to interpret multi-wavelength photometry~\citep{Bruzual_2003,Conroy_2013}. Over the past two decades, several powerful SED fitting tools have been developed and widely adopted, e.g., LePhare~\citep{Arnouts_2002,Ilbert_2006}, PROSPECTOR~\citep{Benjamin_2021}, CIGALE~\citep{Boquien_2019}, MAGPHYS~\citep{da_Cunha_2008}, EAZY~\citep{Brammer_2008} and BAGPIPES~\citep{Carnall_2018}. Among all, stellar mass estimates rely primarily on the integrated stellar luminosity, with low- to intermediate-mass stars contributing significantly in the rest-frame optical and near-infrared (NIR) bands. These stars, being less affected by extinction, enable relatively robust ${M_\star}$ estimates with an assumed initial mass function~\citep[IMF; e.g.][]{Kroupa_2001} \citep{Salvato_2018}.
SFR is primarily inferred from the radiation of young, massive stars, which emit strongly in the ultraviolet (UV) and whose emission is often absorbed by dust and re-emitted in the infrared (IR). To break the degeneracy between dust attenuation and stellar population age, SFR estimates typically combine UV and far-infrared (FIR) luminosities~\citep[e.g.][]{Pannella_2009,Buat_2010,Hao_2011,Riccio_2021}. Although effective, SED fitting is computationally expensive and sensitive to model assumptions regarding, e.g., stellar evolution, star formation histories and dust attenuation, particularly in the absence of FIR constraints~\citep[e.g.][]{Michalowski_2014, Wuyts_2011}.

In the coming decade, the launch of next-generation space-based telescopes and the ongoing accumulation of survey data from missions will statistically enhance our understanding of galaxy evolution. Among these, the China Space Station Telescope \citep[CSST; ][]{Csstcollaboration2025} and \textit{Euclid}~\citep{Euclid} hold significant potential for these large-sample studies. The CSST aims to take the survey camera roughly 7 years of operation accumulated over 10 years of orbital time to image roughly 17,500 $\rm deg^2$ (roughly 300 times wider view than the Hubble Space Telescope (HST)) of the sky in NUV, $u,\ g,\ r,\ i,\ z,\ \text{and}\ y$ bands~\citep{Zhan2021}, enabling accurate photo-$z$ estimation. It can obtain more than one billion galaxy images and one hundred million galaxy spectra, and discover millions of active galactic nuclei (AGN) and other astronomical objects in a large redshift range. When combined with \textit{Euclid}’s three NIR bands of $Y,\ J,\ \text{and}\ H$~\citep{Laureijs2011}, one can obtain robust estimates of $M_\star$.
However, the influx of billions of sources will further exacerbate the computational burden of traditional template-fitting methods. Moreover, the traditional SED-fitting method is sub-optimal for constraining SFR in the absence of MIR and FIR data~\citep{Pannella_2009,Buat_2010,Riccio_2021} which both instruments lack, unless supported by external constraints or informed priors that help to break the dust–age degeneracy~\citep[e.g.][]{Arango_Toro_2025}. Therefore, the challenge behind the large galaxy sample that the CSST will provide lies in, how to efficiently process billions of galaxies, how to robustly constrain the SFRs of large galaxy populations using only limited optical and NIR bands, and, more crucially, whether it is feasible to effectively leverage JWST data to calibrate other surveys.

In this context, machine learning (ML) techniques have emerged as powerful tools for analyzing complex, high-dimensional astronomical datasets \citep[for a review, see e.g.][]{Ball_2010, Baron_2019, Longo_2019, Salvato_2018}.
Among these techniques, Self-Organizing Maps \citep[SOMs;][]{Kohonen_1982} display distinct advantages in handling large astronomical datasets and visualizing high-dimensional data. SOM is an unsupervised learning algorithm that allows us to project high-dimensional data onto a two-dimensional map while preserving topological structure, making it particularly useful for analyzing large astronomical surveys. It can be employed as a computationally efficient alternative to traditional SED fitting, capable of processing vast amounts of data and exploring complex, non-linear parameter spaces. Previous studies have successfully applied SOMs for tasks such as classifying stellar spectra \citep[e.g.][]{Mahdi_2011}, classifying galaxy morphology \citep[e.g.][]{Galvin_2019}, estimating photometric redshifts and other parameters \citep[e.g.][]{Masters_2015, Hemmati_2019, Davidzon_2019, Latorre_2024}, as well as selecting specific targets \citep[e.g.][]{Masters_2015, Hemmati_2019}. \citet{Davidzon_2022} successfully applied this approach to physical parameter estimation using the COSMOS2020 dataset \citep{Weaver_2022}. However, traditional SOM algorithms face challenges in handling missing values and require strict consistency between input color combinations and the training set, imposing stringent requirements on the input data.

In this study, we aim to prepare for upcoming CSST surveys and fully exploit the potential of JWST’s comprehensive and high-quality data. We introduce a hybrid method that combines SOM with SED fitting techniques (SHAPE, SOM-SED Hybrid Approach for efficient Parameter Estimation), utilizing JWST’s high-precision photometric data to establish a robust reference sample for future large-scale surveys. As a test of the methodology, we train a SOM using the deep data from the JWST PRIMER survey~\citep{Primer}, preliminarily focusing on galaxies at redshifts $z \sim 1.5 - 2.5$. SOM is employed to cluster galaxy samples and derive the average SED for each cell, thereby establishing an SED library (SED Lib hereafter) within the given redshift range. For any given galaxy, its colors are compared with the SEDs in the SED Lib to obtain the probability distribution over the SOM, followed by the corresponding parameter estimates. SHAPE extends the SOM-based approach from handling fixed discrete photometric points to a continuous functional framework and has been demonstrated to be efficient when applied to the COSMOS2020 catalog, which we employ as the test dataset. The basic workflow is described in Fig.~\ref{fig:process}, and more details are discussed in Sect.~\ref{sec:methods}. 

This paper is structured as follows. The employed JWST and COSMOS2020 photometric catalogs, along with their physical parameters, are summarized in Sect.~\ref{sec:data}. 
In Sect.~\ref{subsec:D21}, we thoroughly describe the mechanisms and setups of SOM, and replicate the parameter estimation method using SOM as presented in \citet{Davidzon_2022}, but with a SOM trained on JWST data. In Sect.~\ref{subsec:SHAPE}, we introduce the SHAPE method, where we detail the construction of the SED Lib, synthesized based on SOM clustering results, and its application to parameter estimation. In Sect.~\ref{sec:results}, we present the parameter estimation results obtained by applying the traditional and hybrid methods to the JWST PRIMER and COSMOS2020 datasets, and compare them with SED-derived estimates. In Sect.~\ref{sec:discussions}, we compare our work with previous work, and discuss the limitations and outlook.
In Sect.~\ref{sec:conclusions}, we present our summary and conclusions.
Throughout this work, we assume the Planck $\LCDM$ cosmology with 
$H_{0}=70\,\kms/\Mpc$, $\Omegam = 0.3$, and $\Omegal = 0.7$. All magnitudes are in the absolute bolometric (AB) system.

\section{Data} \label{sec:data}

\begin{figure*}
    \centering
    \includegraphics[width=0.8\linewidth]{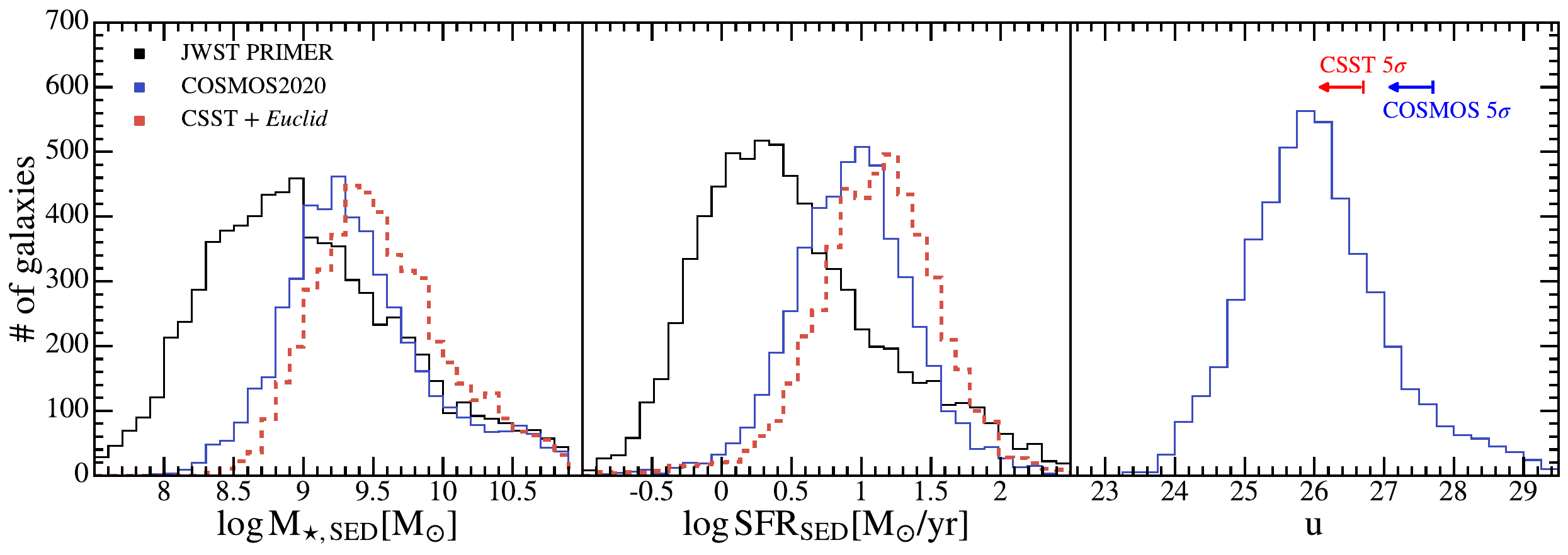}
    \caption{Observational properties of the galaxy samples used throughout this work, selected according to the criteria in Sect.~\ref{subsec:sel}. \textit{Left}: Stellar mass distribution of the JWST training set (black), the COSMOS2020 test sample (blue) and the mock CSST+\textit{Euclid} catalog (red dashed). \textit{Middle}: SFR distribution of the three samples. \textit{Right:} The magnitude distribution of COSMOS2020 sample, where we show the depths at $5\sigma$ for COSMOS2020 and CSST respectively.}
    \label{fig:data}
\end{figure*}

To train a SOM that is both accurate and comprehensive in covering a wide variety of galaxies, we employ the PRIMER survey, which, at the time of writing, is the largest publicly available JWST survey. It provides extensive and deep NIR photometry as well as substantial MIRI coverage, making it an ideal choice for constructing the training dataset in our work. In addition, COSMOS2020 galaxies are used as a test sample to validate our methodology and assess its reliability.

\subsection{JWST PRIMER photometric catalog}\label{subsec:jwst}

In this study, we utilize one of the largest and deepest JWST/NIRCam and JWST/MIRI surveys from public Treasury Programs, the Public Release IMaging for Extragalactic Research \citep[PRIMER, GO 1837, ][]{Primer} program. The PRIMER survey targets the CANDELS-COSMOS and CANDELS-UDS fields with deep imaging in 10 bands: F090W, F115W, F150W, F200W,
F277W, F356W, F444W and F410M with NIRCam, and F770W and F1800W with MIRI.
All the JWST/NIRCam and previous HST images are publicly available from the Grizli Image Release v7.0 1~\footnote{\url{https://dawn-cph.github.io/dja/imaging/v7}}, which were reduced and processed by the Grizli pipeline \citep{brammer_2022_grizli}. Due to a few issues in the default MIRI pipeline, the F770W and F1800W data were reprocessed using a custom-made pipeline to reconstruct the multiwavelength source catalog. The detailed technics are discussed in \cite{Wang_2025}.

As the reference sample for SOM training and labeling, JWST galaxies are required to have not only broadband colors but also measurements of key physical parameters. We derive photometric redshifts using EAZY \citep{Brammer_2008}, followed by the estimation of rest-frame colors, stellar mass and 100 Myr time-averaged SFR using BAGPIPES \citep{Carnall_2018}. Here in summary, we adopt the delayed star formation history with the e-folding timescale between [0.01, 10]~Gyr, the Calzetti extinction law~\citep{Calzetti+2000} for young and old stellar populations separately with $A_{\rm V}$ between [0, 5], nebular emission in BAGPIPES produced by CLOUDY~\citep{Ferland+2017,Byler+2017,Carnall_2018} with $\log\ U$ between [-5, -2], stellar metallicity between [0.01, 2.5]~$Z_\odot$, and the time since star formation begins between [0.03, 10]~Gyr. Further details on catalog construction and parameter estimation are provided in \cite{Wang_2025}.

\subsection{COSMOS2020 photometric catalog}\label{subsec:cosmos2020}
Since one of our goals is to bridge two different photometric systems using SOM, thereby leveraging JWST data to calibrate the physical parameters estimated from other photometric surveys, we employ the COSMOS2020 photometric catalog as the test sample and the basis of the mock CSST and \textit{Euclid} catalog.

COSMOS2020 \citep{Weaver_2022} is an updated version of the previous COSMOS2015 catalog \citep{Laigle_2016}. Source detection and multi-wavelength photometry is performed for 1.7 million sources across the $2\,\rm deg^2$ of the COSMOS field, $\sim966,000$ of which are measured with all available broad-band data, typically in twelve bands ($u,\ g,\ r,\ i,\ z,\ y\, \text{with MegaCam/CFHT},\ Y,\ J,\ H,\ K_{\rm s}$ with VISTA/VIRCAM, and ch1, ch2 with \textit{Spitzer}/IRAC). In addition, most sources are provided with ancillary medium-band and narrow-band photometry, e.g. NB118 with \textit{VISTA}/VIRCAM, and IB427, IB464, IA484, IB505, IA527, IB574, IA624, IA679, IB709, IA738, IA767, IB827, NB711, NB816 with \textit{Subaru}/Suprime-Cam.
We incorporate a well-matched joint catalog~\citep{Wang_2024_cosmos} as ancillary data which extends the photometry to \textit{Spitzer}/MIPS (24 $\rm{\mu m}$), the Herschel PACS (100 and 160 $\rm{\mu m}$) and SPIRE wavebands (250, 350 and 500 $\rm{\mu m}$), further improving the robustness of parameter estimates.

Rather than utilizing the physical parameters derived from SED fitting with LePhare \citep{Arnouts_2002,Ilbert_2006} in the original catalog, following the procedure in Sect.~\ref{subsec:jwst} and \cite{Wang_2025}, we derive new parameters based on a subset of at least 12 broadbands, with a total of up to 33 bands. This approach ensures consistency by mitigating discrepancies introduced by different codes and templates in the comparison. Nevertheless, our estimates show excellent agreement with those provided in the COSMOS2020 public catalog.

\subsection{The mock CSST and \textit{Euclid} catalog}\label{subsec:csst}
CSST \citep{CSST1,CSST2,Csstcollaboration2025} is a major science project initiated by the China Manned Space Program and is planned to be launched around 2026. The imaging bands consist of NUV, $u,\ g,\ r,\ i,\ z,\ \text{and}\ y$, covering a wide wavelength range from 2000 {\AA} to 1.1 $\rm{\mu m}$ with a pixel scale of 0".074. The slitless spectroscopy bands include GU ($255-400$ nm), GV ($400-620$ nm), and GI ($620-1000$ nm) with the same pixel scale. The planned surveys for both imaging and spectroscopy include (but are not limited to) a wide-field survey over about 17,500 $\rm deg^2$ sky area and a deep-field survey of about 400 $\rm deg^2$. The wide-field imaging survey will reach an average limiting magnitude better than 25.5 mag at $5\sigma$ for point sources.

\textit{Euclid} \citep{Euclid_2025} is an ESA M-class astrophysics and cosmology mission. It's equipped with two instruments, i.e. the optical VIS imager and the Near-Infrared Spectrograph and Photometer (NISP).
The VIS imager comprises 36 4k × 4k CCDs with 0".1/pixel, targeting at a broad 550–900 nm optical bandpass and reaching $\sim24.5$ AB mag at 10$\sigma$ for galaxies with a size larger than 1.25 times the full width at half-maximum (FWHM) of the PSF~\citep{VIS_2025}. The NISP includes 16 arrays of 2k×2k NIR-sensitive HgCdTe detectors with a pixel scale of 0".3, and will obtain NIR imaging in Y, J, and H bands for photometric redshifts reaching 24 mag at 5$\sigma$ for point sources in the Wide Survey~\citep{NISP_2025}.
It will survey 15,000 $\rm deg^2$ of the extragalactic sky avoiding the ecliptic plane due to increased zodiacal background in its Wide Survey, plus 40 $\rm deg^2$ split over three deep fields with a depth increase by 2 mag. We refer to \citet{Liu_2023} for further details on the scientific synergies between CSST and \textit{Euclid}.

To simulate the application of our model on CSST data, we select photometry from the COSMOS2020 catalog that corresponds to the nine bands available in the CSST and \textit{Euclid}, i.e. $u,\ g,\ r,\ i,\ z,\ y,\ Y,\ J,\ \text{and}\ H$, and then derive an additional set of physical parameters with the nine bands, adopting the same methodology. This set of parameters only serves as a baseline for comparison with the full-band photometry, allowing us to assess the extent to which SOM- and SED-based estimates can approximate the "true values" derived from comprehensive photometric datasets. 
Here we focus solely on evaluating the performance of parameter estimation using photometry from the selected filters, without strictly requiring the source selection to match the target population of CSST observations. Nonetheless, we note that the majority of COSMOS2020 galaxies fall within the observable range of the CSST deep survey (see Fig.~\ref{fig:data}).

\subsection{Selection functions}\label{subsec:sel}
The JWST catalog is limited to sources with $\rm S/N_{\rm F444W} > 7$ and requires non-zero detections in all nine bands. The former criterion is adopted empirically to exclude spurious detections, while the latter is imposed due to the current model’s inability to handle missing data (see Sect.~\ref{subsec:limit}). We also exclude objects with $ \rm S/N_{\rm F770W} < 2$ which is crucial in determining stellar mass and constraining SFR~\citep[][]{Wang_2025}. Additionally, we constrain the training sample to the redshift range  $1.5<z<2.5$, as there is a lack of rest-frame UV-optical bands for lower-redshift galaxies, which are essential for reliable SFR estimation. Notably, we did not incorporate HST bands to train our SOM, as the photometric depth differences and measurement uncertainties between the HST and JWST could potentially affect clustering results. However, we did include them during SED fitting to derive accurate physical properties. In principle, it would be feasible to extend the sample coverage to lower redshifts using available optical and UV bands and to higher redshifts with larger fields and more samples (see Sect.~\ref{subsec:train}). After applying the selection criteria, the JWST training sample consists of 7,507 galaxies.

For the COSMOS2020 catalog and the mock catalog, we apply a $3\sigma$ cut, corresponding to a $K_s<24.2$ and require non-zero detection in all twelve broadband filters. To maintain structural consistency between the two datasets, we limit the sample to the redshift range $1.5<z<2.5\, (\pm0.1)$. A random subset of 5,000 sources is extracted from the selected sample as the test set and the basis of the mock catalog.
For the mock catalog, despite the differences in observational depths between CSST and COSMOS2020 due to instrument variations, the majority of our sample lies within the $5\sigma$ detection limit of the \textit{u}-band for CSST’s deep field survey. Therefore, in this study, the mock catalog and the COSMOS2020 catalog share the same set of sources, differing only in the selection of filters used during SED fitting.
Unless explicitly stated, all parameter estimates from COSMOS2020 are derived from the full-band photometry (the "true values"), which serves as the reference for comparison with our SOM-based estimates.

In Fig.~\ref{fig:data}, we show the parameter distribution of galaxy samples used throughout this work. In the left and middle panels, we show that the composition of the observational samples from JWST and COSMOS2020 differs in parameter space. The latter tends to favor high-$M_\star$, high-SFR galaxies, with median values exceeding those of the JWST sample by 0.3 dex in $M_\star$ and 0.6 dex in SFR. However, overall, its distribution range is well represented by the JWST sample. In the absence of $K_{\rm s}$, ch1 and ch2 bands, the SED fitting results for $M_\star$ and SFR from the mock CSST+\textit{Euclid} catalog show some degree of overestimation, which we will discuss in Sect.~\ref{sec:vs}.


\section{Methods}\label{sec:methods}
\subsection{Parameter estimation in traditional SOM-based method}\label{subsec:D21}
\subsubsection{Self-organizing maps}\label{subsec:som}
Self-organizing maps (SOM; \citealt{Kohonen_1982}) are employed for classifying and analyzing multi-dimensional galaxy photometric data. SOM-based parameter estimation involves two main steps: unsupervised classification and supervised labeling. Below, we outline the algorithm used during the unsupervised phase.
In the unsupervised training phase, the SOM is governed by a well-defined mathematical framework consisting of initialization, competition, cooperation, adaptation, and iteration.
First, a two-dimensional grid of nodes is constructed, each associated with a prototype vector of galaxy colors. These prototype vectors $\vec{w}_j = (w_{1,\,j}, w_{2,\,j}, \dots, w_{{N_{\rm dim}},\,j})$, hereafter referred to as “weights”, are initialized using principal component analysis (PCA) and iteratively refined through a competitive learning process.
During each iteration, an input vector $\vec{c} = (c_1, c_2, \dots, c_{N_{\rm dim}})$ is presented, and each neuron computes its discriminant function to measure similarity via
\begin{equation}
   \Delta = \sqrt{\sum_{N_{\rm dim}} (c_i-w_i)^2}\,,
   \label{eq:distance}
\end{equation}
where the neuron with the smallest $\Delta$ is identified as the best matching unit (BMU).

Following the competition step, a cooperation phase is introduced. In this phase, not only the best matching unit (BMU) but also its neighboring neurons in the two-dimensional lattice are updated to reduce their distance from the input vector to better fit the training data. The degree of adjustment decreases with both spatial distance from the BMU and iteration time, following a Gaussian function where the neighborhood radius $T$ shrinks exponentially:

\begin{equation}
   T_{j,I(c)}(t) = \exp \left( -\frac{S_{j,I(c)}^2}{2\sigma(t)^2} \right),
   \label{eq:ite}
\end{equation}
where $S_{j,I(c)}$ is the distance between neuron $j$ and the BMU $I(c)$ of input $\vec{c}$. The learning rate $\sigma(t) = \sigma_0 \exp(-t/\tau_{\sigma})$ decreases over time, where $t$ denote the current number of iterations, $\tau_{\sigma}$ and $\sigma_{0}$ are free parameters controlling the rate of decay. This ensures that distant nodes are updated less significantly. These updates progressively align the SOM with the underlying structure of the input data, resulting in a topologically ordered grid where similar galaxy colors are expected to be clustered together. 

Finally, the process iterates up to the maximum number of iterations, during which the learning rate gradually approaches zero, ultimately yielding a topologically ordered and stable map. Galaxies associated with the same BMU are expected to exhibit similar SEDs, modulo their brightness (i.e., a normalization factor) and the scatter introduced by photometric uncertainties. In this work, we utilize the Python library miniSOM~\citep{minisom} for SOM construction and training.

\subsubsection{SOM training}\label{subsec:train}

\begin{figure}
    \centering
    \includegraphics[width=0.95\linewidth]{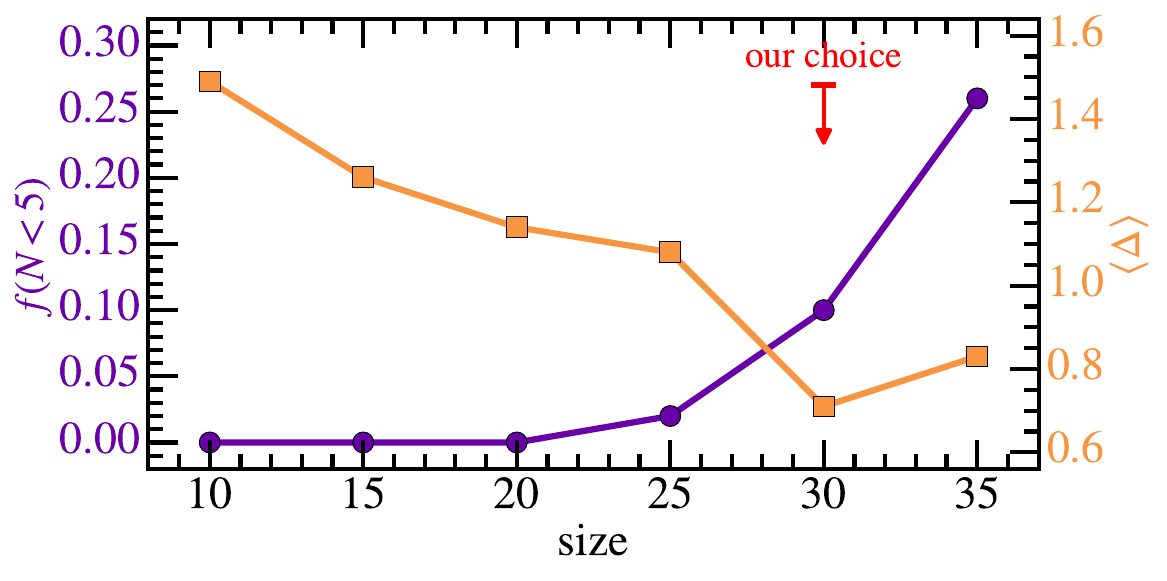}
    \caption{Undersampling fraction (the fraction of SOM cells containing fewer than five galaxies; purple) and quantization error (the average $\Delta$ in Eq.~\ref{eq:distance}; orange) as a function of SOM size, given a limited data volume of 7,507 galaxies. While the undersampling fraction increases monotonically with size, the quantization error first decreases and then increases, reaching an optimal value at a SOM size of $30\times30$.}
    \label{fig:size_test}
\end{figure}

In this work, the input parameter space consists of eight dimensions, with each representing a observed-frame adjacent color of galaxies. These colors, derived from the photometry of the JWST PRIMER catalog (see Sect.~\ref{subsec:jwst}), are: F090W-F115W, F115W-F150W, F150W-F200W, F200W-F277W, F277W-F356W, F356W-F410W, F410W-F444W, and F444W-F770W. The colors are normalized to have zero mean and unit variance, ensuring that each color contributes equally during training, independent of its intrinsic range or scatter. 

The performance of SOM depends on the user-defined map size and geometry, which should be chosen to balance adequate data sampling with sufficient resolution (i.e., minimizing quantization error). For the SOM geometry, \citet{Davidzon_2019} demonstrated that a square topology yields the most effective parameter estimates, and we follow their recommendation accordingly. To determine the optimal map size, we start with a $10 \times 10$ grid and gradually increase the size in increments of five. For each configuration, we assess the quantization error (i.e. the average $\Delta$ of the SOM) and the fraction of the undersampling cells (i.e. the cells that associate to fewer than five galaxies). In general, the average number of galaxies per cell decreases with increasing SOM size due to the limited data volume, whereas the quantization error decreases initially but rises again beyond a map size of 30 (see Fig.~\ref{fig:size_test}). Therefore, we adopt a relatively moderate SOM size of 30×30, smaller than those used in \citet{Davidzon_2022, Latorre_2024}, to maintain a reasonable trade-off between parameter resolution and robust sampling.

\begin{figure*}
    \centering
    \includegraphics[width=0.25\linewidth]{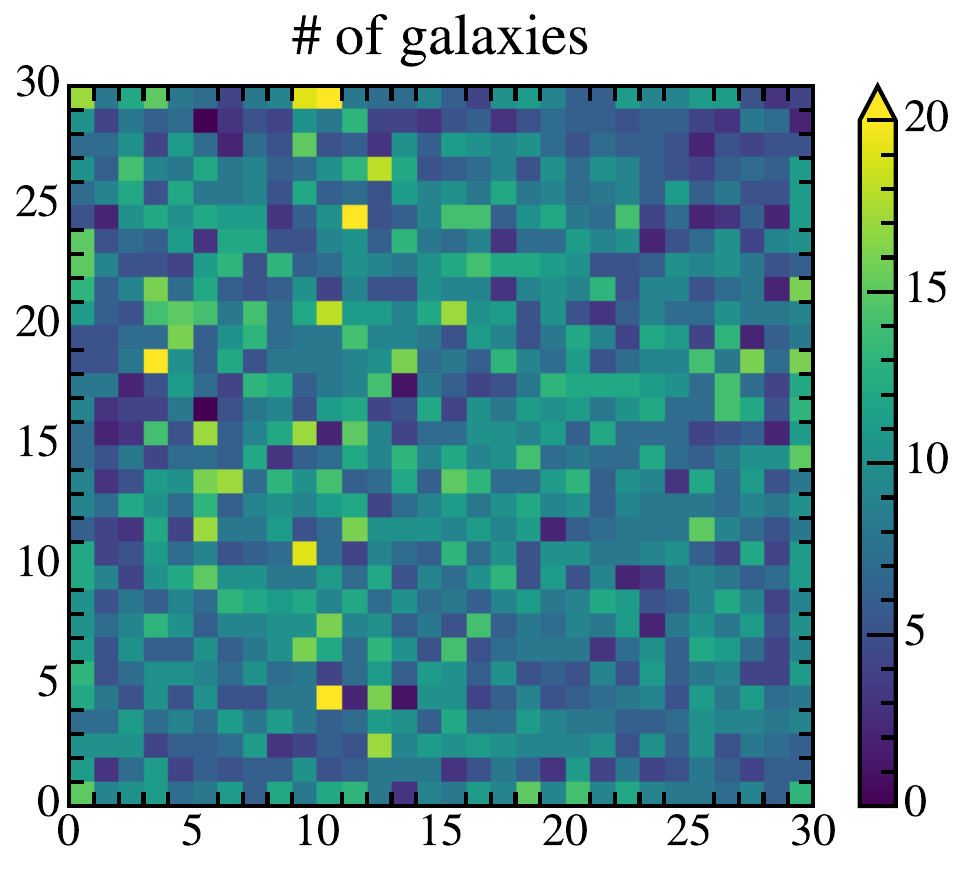}
    \includegraphics[width=0.25\linewidth]{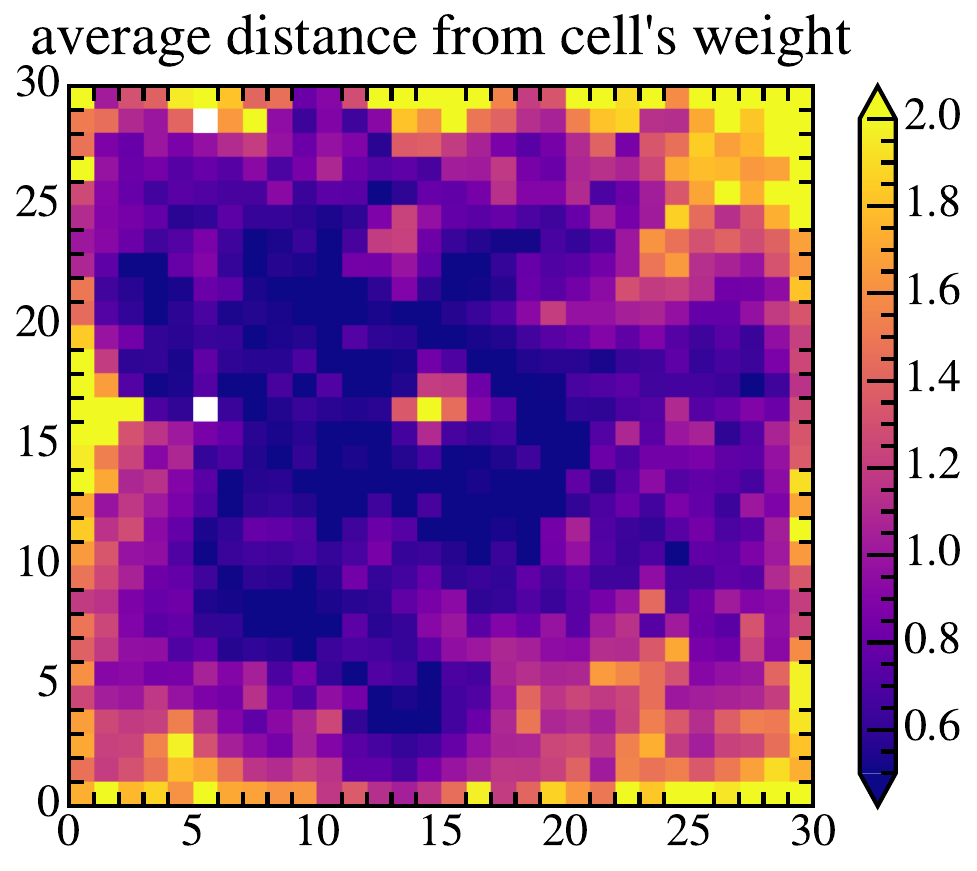}
    \includegraphics[width=0.25\linewidth]{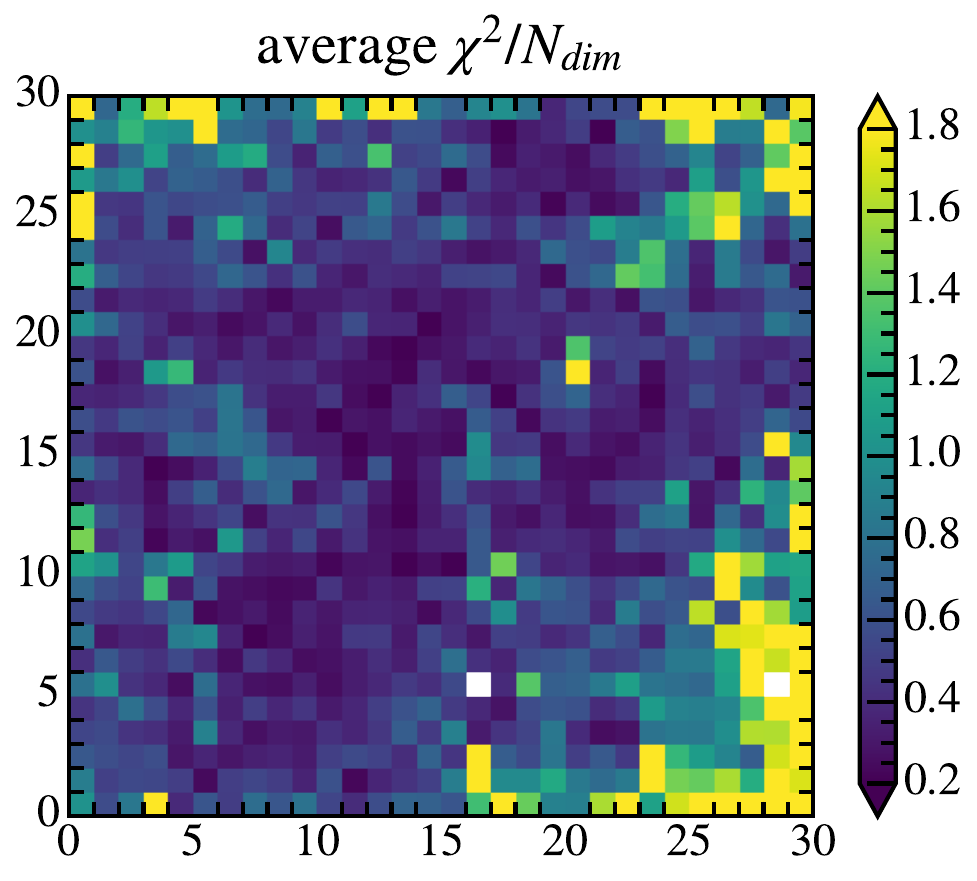}
    \caption{SOM of JWST PRIMER galaxies at $z\sim 1.5-2.5$ selected as described in Sect~\ref{subsec:sel}. \textit{Left:} Number of galaxies per cell. \textit{Middle:} Similarity between galaxies in a given cell and the corresponding SOM weight, quantified using Eq~\ref{eq:distance}. \textit{Right:} Similarity incorporating photometric errors. Compared to \citet{Davidzon_2022}, the scatter is significantly reduced when applying the JWST catalog.}
    \label{fig:som}
\end{figure*}

Our final 30×30 SOM is shown in Fig.~\ref{fig:som}. The galaxy distribution across the map is relatively uniform, with 90\% of cells containing statistically sufficient samples (more than five galaxies), and only two cells being empty (see Fig.~\ref{fig:som}, left panel). To evaluate the accuracy of the SOM in representing the JWST data manifold, we compute the Euclidean distance between each galaxy and its corresponding cell weight using Eq.~\ref{eq:distance}.
Across the cell the average $\Delta$ per cell is $< 1$ (Fig.~\ref{fig:som}, middle panel).

For marginal cells with relatively high $\Delta$ values, we find that they predominantly contain extreme or low S/N galaxies. This phenomenon is consistent with the boundary effect observed in previous studies \citep{Davidzon_2019}, where irregular samples tend to populate the outermost regions of the map. An additional metric to evaluate clustering performance, incorporating photometric uncertainties, is the reduced $\chi^2$ distance, defined as:

\begin{equation}
   \chi^2_{\rm SOM} = \sum_{N_{\rm dim}} \frac{(c_i-w_i)^2}{\sigma_i^2}
   \label{eq:chi2}
\end{equation}
where $\sigma_i$ is the normalized photometric error of the $i$ th color. Despite the presence of large uncertainties concentrated near the edges, our analysis indicates significantly lower $\chi^2$ values compared to \citet{Davidzon_2022}. This demonstrates that, despite the smaller sample size, reduced SOM dimensions, and fewer colors, the precise photometry and broad wavelength coverage of JWST substantially enhance the clustering accuracy of SOM. As shown in Sect~\ref{subsec:label}, regions with high quantization error, primarily located in the lower-right corner of the SOM, correspond to massive, older, and quiescent galaxies. The clustering of these galaxies is significantly influenced by the diversity of dust emission in the mid-infrared, which, given the current JWST sample size, cannot yet be resolved in fine detail.

We further verify that the distributions of the eight colors in the input dataset are consistent across SOM cells, ensuring that the SOM accurately represents the input data. Fig.~\ref{fig:colors} shows the distribution of the eight weight dimensions (representing the normalized colors) across the map. The deviation between the median colors and the SOM vectors within each cell (see Fig.~\ref{fig:dev}) is defined as the difference between the two, normalized by the scatter of the respective color. Across all cells, the deviations remain significantly smaller than 1 $\sigma$, with no notable differences among colors, confirming that each color contributes equally to the classification.

\subsubsection{SOM pixel labeling}\label{subsec:label}
Galaxies with similar colors assigned to the same or neighboring cells are expected to exhibit similar physical properties, e.g. $M_\star$ and SFR which can be used to characterize the properties of each cell (hereafter referred to as the “label”). Unlike \citet{Davidzon_2019} that adopt the median value for labeling, we label SOM pixels using the mode of each parameter distribution (i.e., the value corresponding to the peak of the distribution within each cell). Since $M_\star$ and SFR vary with the amplitude of a galaxy’s spectrum, which is not explicitly modeled in our color-trained SOM, we normalize these values by rescaling them to a reference magnitude of $\rm F444W=26$ via

\begin{equation}
\rm nX=X \times 10^{-0.4\times(26-F444W)},
\label{eq:norm}
\end{equation}
where X is the the galaxy parameter to be normalized, and F444W is the F444W-magnitude for each individual galaxy. The F444W band is chosen for normalization throughout the analysis, as for galaxies in the redshift range $1.5<z<2.5$, it contains fewer emission lines and is predominantly continuum-dominated. Additionally, the PRIMER survey provides the deepest imaging in this band, yielding the highest S/N ratio for sources.

\begin{figure*}
    \centering
    \includegraphics[width=0.475\linewidth]{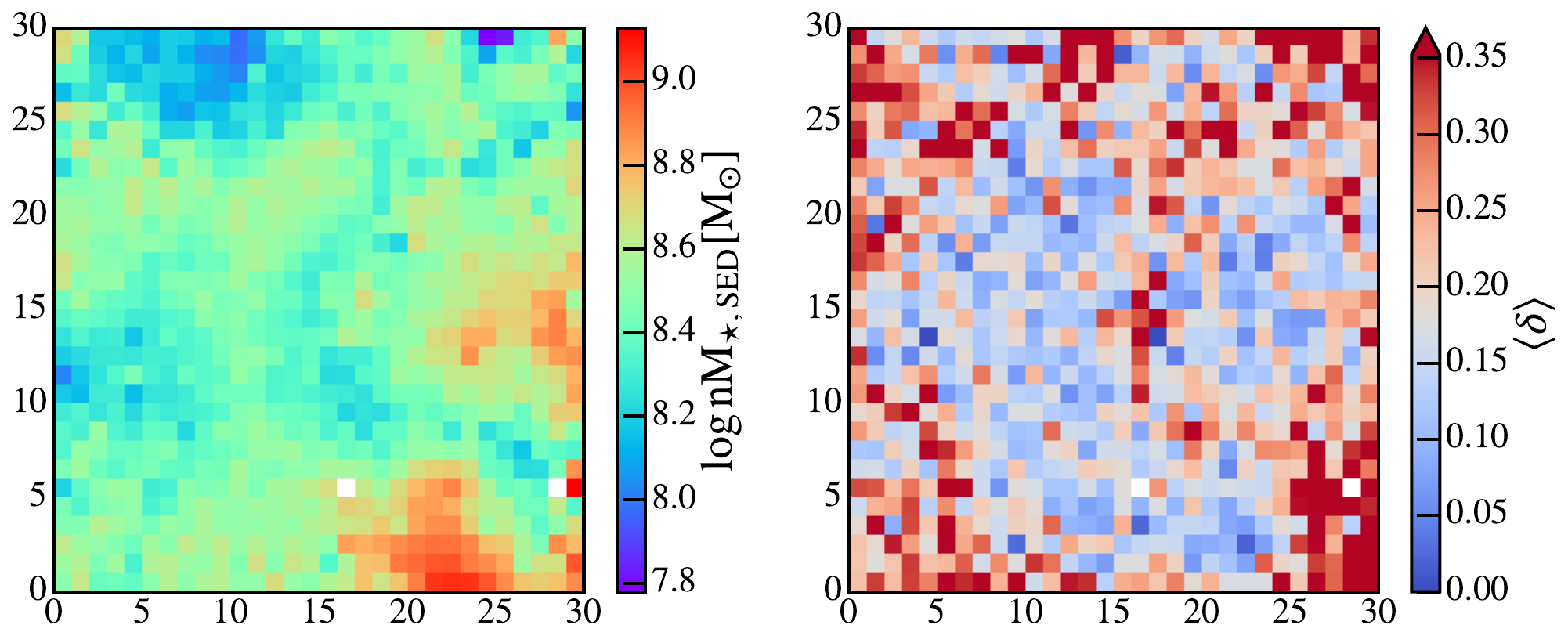}
    \includegraphics[width=0.475\linewidth]{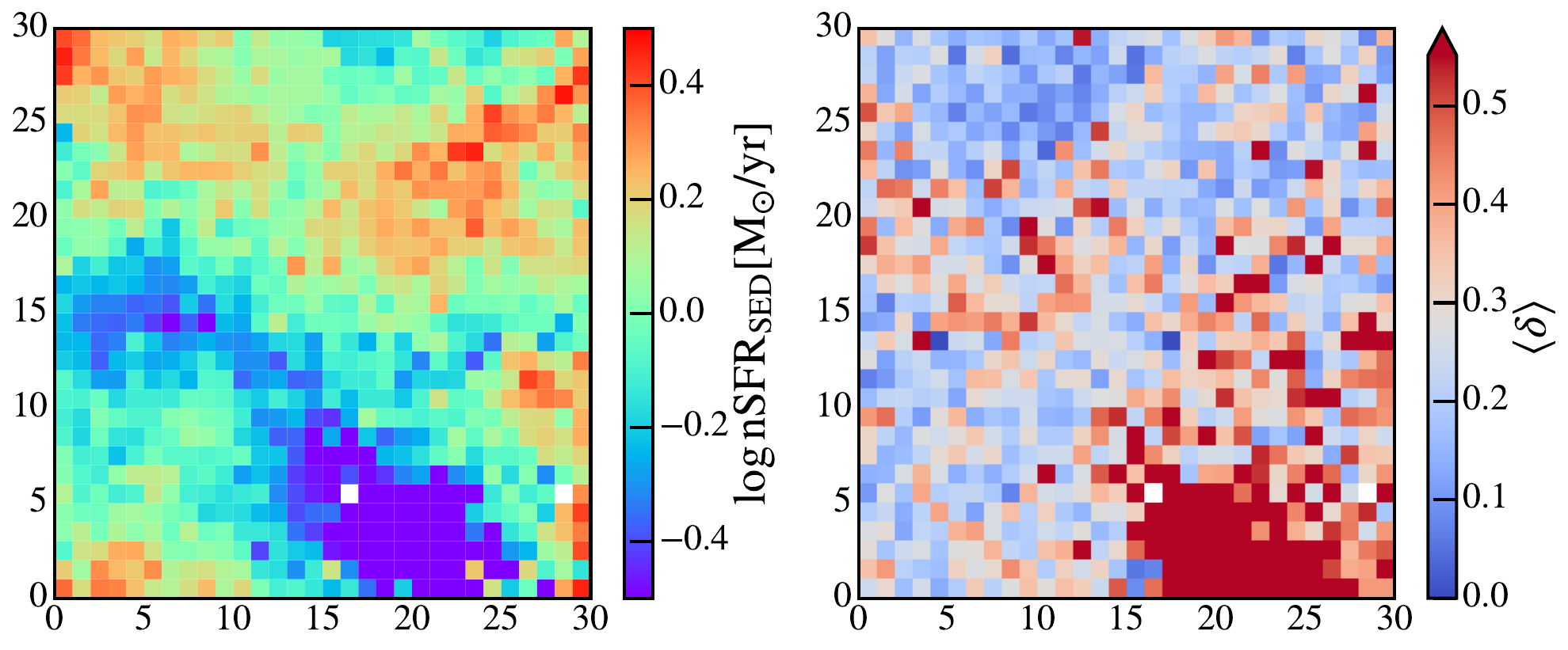}
    \caption{Label maps of normalized stellar mass and SFR. For each pair, the left panel displays the SOM grid color-coded by the mode of the corresponding parameter (stellar mass or SFR), while the right panel shows the distribution width, defined as the 84th–16th percentile range $\langle\delta\rangle = \langle84\%-16\%\rangle$, within each cell.}
    \label{fig:labels}
\end{figure*}

The left panels of Fig.~\ref{fig:labels} show the distribution of normalized stellar mass (${\rm n}M_\star$) and SFR ($\rm nSFR$) across the SOM. We also check the distribution width $\delta$ for each parameter as shown in the right panels of Fig.~\ref{fig:labels}. The width is quantified as the difference between the 84th and 16th percentiles the galaxy properties within each pixel. The $\delta/2$ represents the systematic uncertainties $\sigma_{sys}$ associated with the redshift and stellar population parameters derived using SOM. Specifically, we set a lower limit of $10^{-1}\ \rm M_\odot/yr$ for nSFR labels. This threshold is motivated by the fact that quiescent galaxies can exhibit arbitrarily low SFRs, causing their labels to deviate significantly from the main sequence and often appear as outliers in logarithmic space.
Including such extreme values in the label maps may bias the weighted averaging process and destabilize subsequent parameter estimation.
Nonetheless, we find that the SOM effectively distinguishes between star-forming and passive populations, and the choice of this lower limit primarily affects only a small fraction ($\lesssim 5\%$) of outlier galaxies in the SFR estimates, with negligible impact on the bulk of the population.

It is evident that the scatter of physical parameters within the cells is generally less than 0.5 dex, with most central cells exhibiting a scatter below 0.2 dex, indicating good clustering performance across the SOM. Comparison with Fig.~\ref{fig:som} reveals that cells with larger scatter are primarily located at the periphery, where observational uncertainties dominate. The most significant scatter is concentrated around low-SFR and high-$M_\star$ cells, where the inherent challenges in estimating SFR make robust measurements impractical and beyond the scope of this study.

\subsubsection{Parameter estimation in the traditional method and its limitation}\label{subsec:test run}

\begin{figure*}
    \centering
    \includegraphics[width=0.275\linewidth]{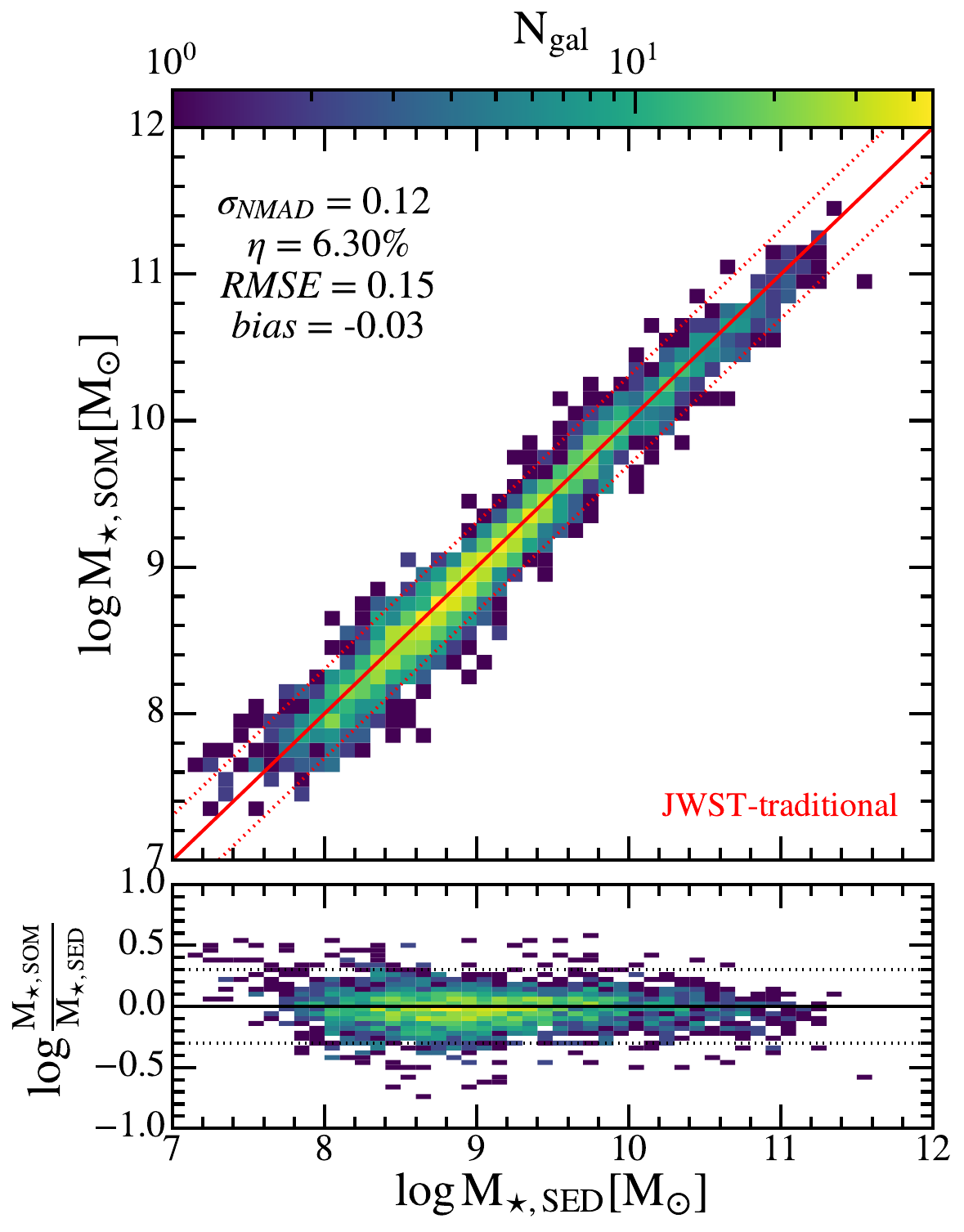}
    \includegraphics[width=0.275\linewidth]{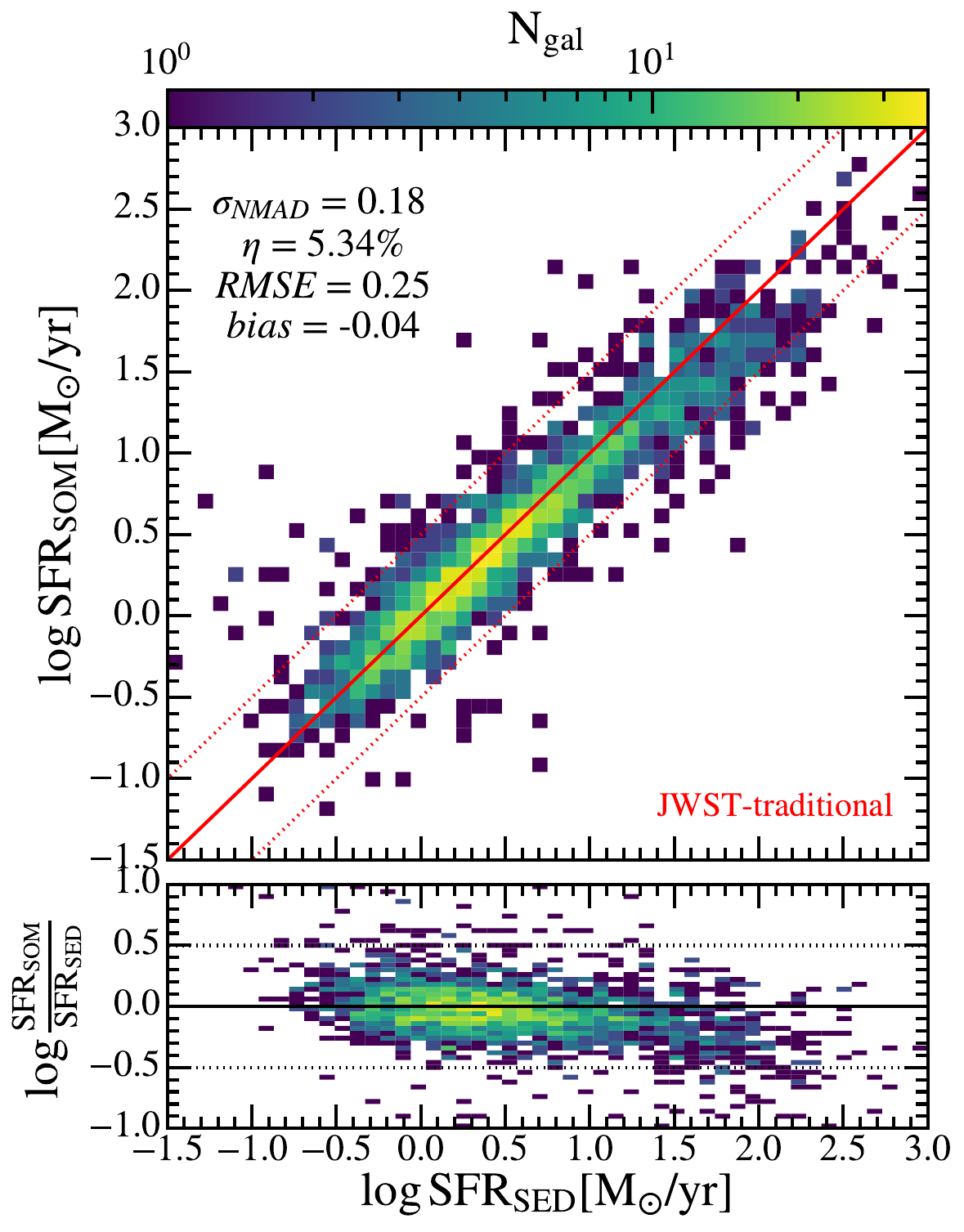}

    \includegraphics[width=0.275\linewidth]{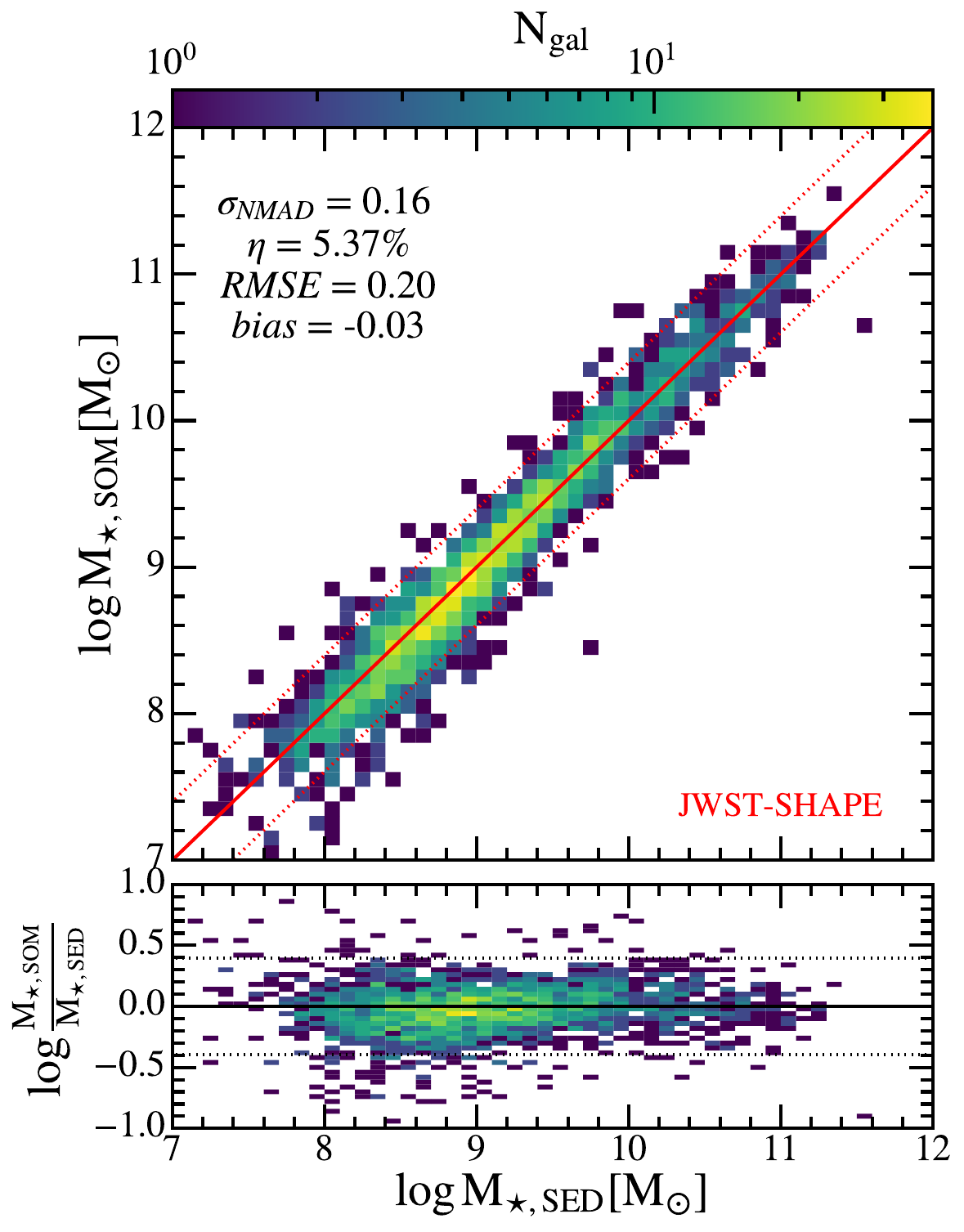}
    \includegraphics[width=0.275\linewidth]{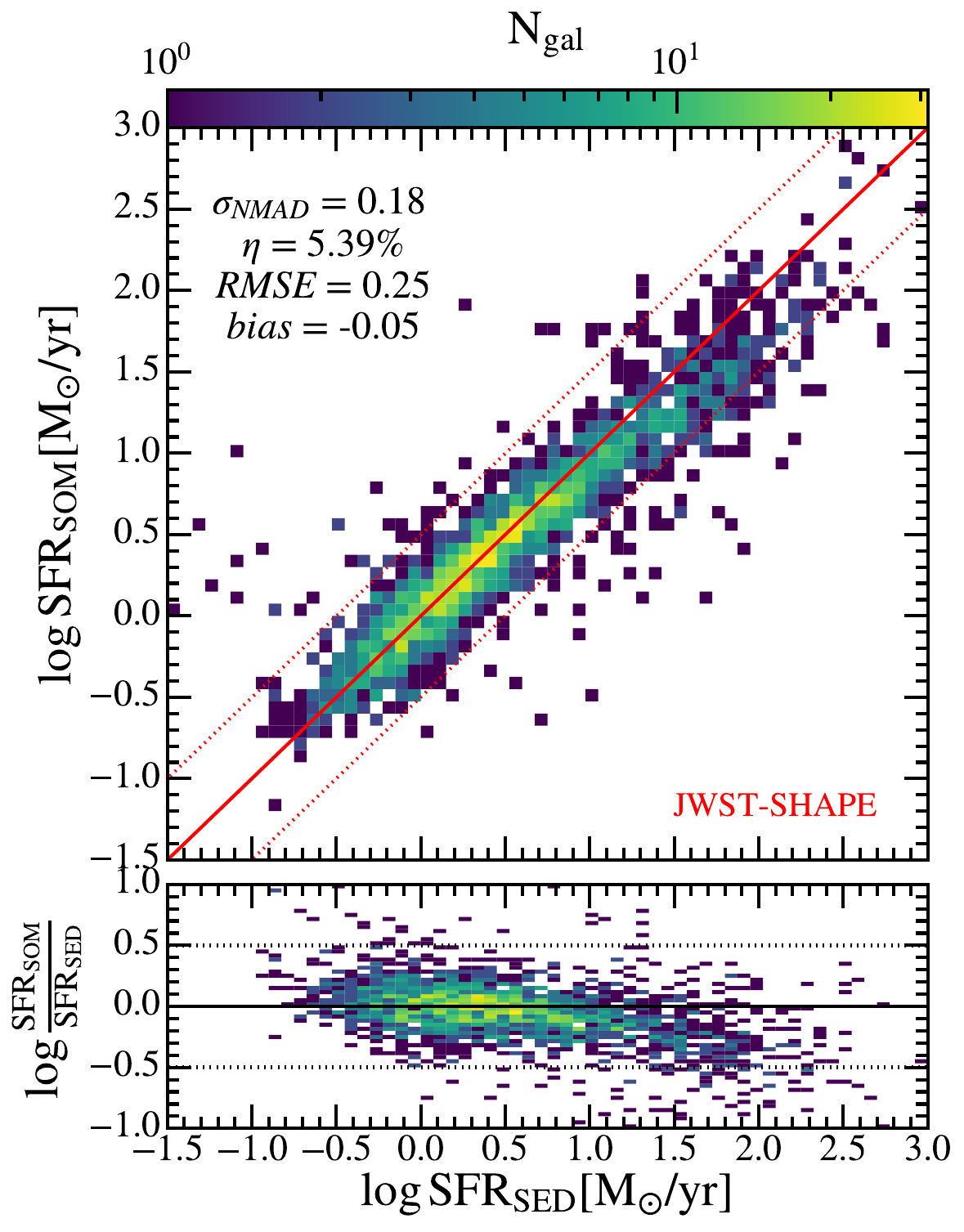}
        
    \includegraphics[width=0.275\linewidth]{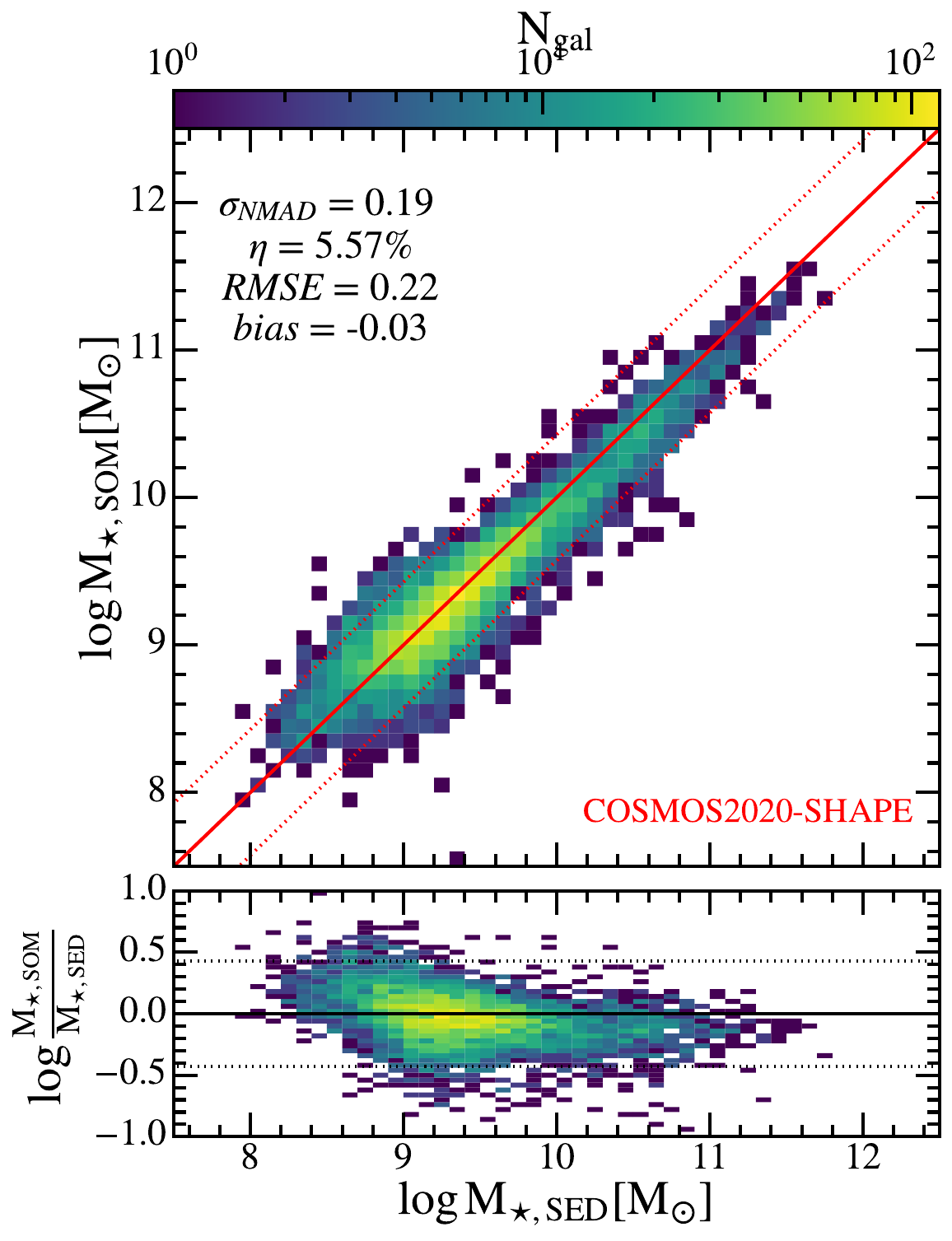}
    \includegraphics[width=0.275\linewidth]{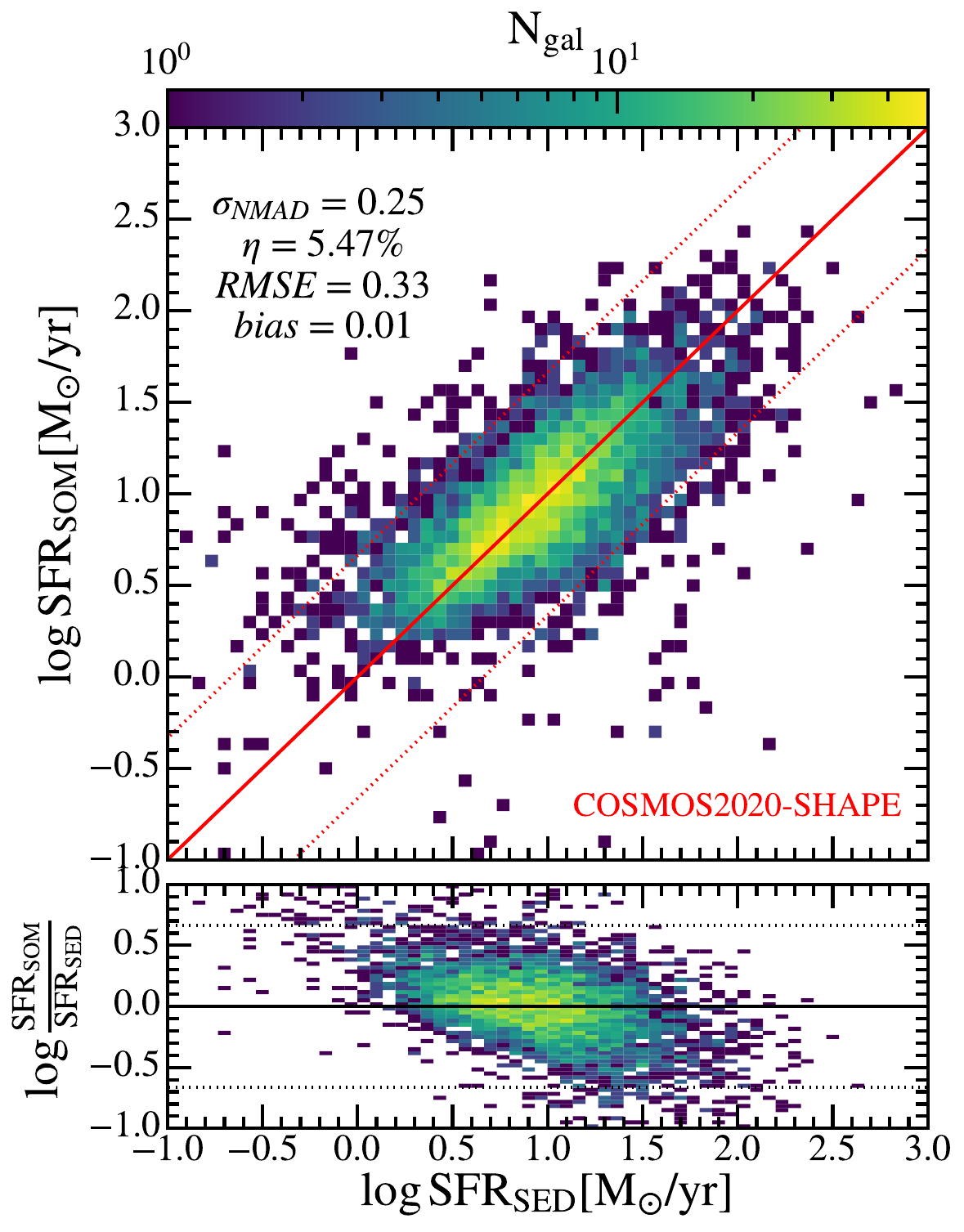}

    \caption{Comparison between stellar mass $M_\star$ and SFR obtained through standard template fitting ($X_{\rm SED}$) and different SOM-based methods presented in this work ($X_{\rm SOM}$, with the specific model highlighted in red in the lower right corner). From top to bottom, the panels show the application of the traditional SOM method to the 7:3 JWST sample, the application of the $25\times25$ hybrid method to the 7:3 JWST sample, and the application of the $30\times30$ hybrid method to the COSMOS2020 sample. The solid red line represents the bisection line, while the red dashed lines indicate the boundaries of the outlier.}
    \label{fig:results}
\end{figure*}

Before applying the method to the COSMOS2020 catalog, we first conduct a sanity check by randomly dividing the JWST sample into two subsets with a 7:3 ratio, designated as the training and test sets, respectively. Following the procedure described above, we trained a 25×25 SOM given the reduced size of the training data. We follow \cite{Davidzon_2022} to estimate $M_\star$ and SFR for the test sample by utilizing the labels of the five cells with the closest weights to each galaxy~\citep{Davidzon_2019, Davidzon_2022}. The parameter estimates were derived from the weighted mean of the target labels and denormalized from the reference magnitude F444W=26, following the equation:

\begin{equation}
    {\rm X_{SOM}} = \frac{\sum_{j=1}^5 \frac{1}{\Delta_j} \rm {nX_{label}}}{\sum_{j=1}^5\frac{1}{\Delta_j}}\times 10^{+0.4\times(26-\rm F444W)}
\end{equation}
where $\Delta_{j}$ is defined in Eq.~\ref{eq:distance} and represents the $j$ th neighbor among the five nearest cells (including the BMU).

The upper panel of Fig.~\ref{fig:results} compares the stellar mass and SFR estimates obtained using SED fitting and the SOM-based method. We also present five statistical metrics for each parameter: normalized median absolute deviation (NMAD), outlier fraction ($\eta$), root mean squared error (RMSE) and bias. The outlier fraction $\eta$ is the percentage of objects satisfying $|\log {\rm X_{SOM}}-\log {\rm X_{SED}}| > 2\sigma$, where $\sigma$ is the standard deviation of the residuals, i.e., log(predicted) - log(true).
It is evident that the parameter estimates derived using the SOM method closely match those obtained through traditional SED fitting, with dispersions of  $\sigma_{\rm NMAD} \sim$ 0.12 and 0.17 for stellar mass and SFR, respectively. Moreover, the computational cost of the SOM approach is reduced to the order of a few CPU minutes, demonstrating its efficiency (see Table~\ref{table:results}).

Despite its computational efficiency and the agreement of its estimates with those from SED fitting, the traditional SOM-based method has notable limitations. It is important to recall that in SOM, galaxy colors are used to generate corresponding weights, with each component representing a specific color. During parameter estimation, the input colors are compared with the weights assigned to each cell. Consequently, the applicability of this method is restricted to galaxy samples that meet the following criteria: (1) they use JWST filters, and (2) all nine bands used in the training set must have detections. As a result, high-precision training results have limited generalizability to other datasets. At the final stages of our work, \citet{Latorre_2024} and \citet{missSOM} independently proposed methods to mitigate the issue of missing values during training and parameter estimation. However, these methods cannot fully address the challenge posed by mismatched filters and thus have not been incorporated into this study. We plan to integrate their approaches in future work to address the missing-data issue during the training phase.

The core limitation of the traditional method lies in its simplification of continuous SEDs into discrete color information. However, for galaxies with similar colors, their SEDs are expected to be correspondingly similar. To overcome this limitation, we extend the discrete model to a continuous framework by fitting SEDs for each cell based on the results of galaxy clustering. This extension allows us to generalize beyond the discrete colors used in the training phase, offering a more flexible and robust approach to parameter estimation.

\subsection{Parameter estimation with SHAPE}\label{subsec:SHAPE}
\subsubsection{Introduction of the hybrid approach of SOM and SED fitting}
Building on the above concept, we introduce SHAPE that utilizes the SOM to cluster galaxies from the JWST catalog and generate representative average SEDs for each SOM cell. In Fig.~\ref{fig:process}, we show a schematic diagram of our model. The model can be roughly divided into three parts, the galaxy automatic classification, construction of the SED Lib and derivation of a new SOM from SED Lib and selective filters. The first part follows the framework of the traditional method, where, if the filters of the test set match those of the training set, galaxies can be directly mapped onto the SOM for parameter estimation (as discussed in Sect.~\ref{subsec:D21}). 

The second part builds upon the galaxy clustering in the first step by incorporating SED templates. From this step onward, the method is no longer purely data-driven, as theoretical models are incorporated. Here we briefly summarize the several advantages of this hybrid method compared to direct SED fitting applied to individual galaxies: (1) By clustering observational data, it eliminates unphysical or unrealistic template combinations; (2) Stacking fluxes enhances both the S/N ratio and wavelength coverage of the photometric points, yielding more accurate SEDs within the SED Lib; (3) The method retains the computational efficiency of the SOM, following the same principles during parameter estimation as in traditional SOM methods.

Once the SED Lib is constructed, each cell corresponds to a normalized SED. In the third part, the filters used in the test set are applied to the SEDs to derive a corresponding set of colors. These colors replace the original SOM weights, forming a new SOM (marked as ${\rm SOM^*}$ hereafter) with components perfectly matching the colors of the test sample. 
This approach allows for flexible selection of photometric bands to be used during the parameter estimation. Below, we provide a detailed explanation of each step in the process.

\subsubsection{Construction of the SED Lib and derivation of the new SOM}\label{subsec:sed}
The construction of the SED Lib follows a series of steps: 
first, we normalize the flux and error in all filters by the flux in the F444W band for each galaxy. Next, we calculate the error-weighted flux ($\bar{f_{\rm i}}$) and error ($\bar{\sigma_{\rm i}}$) in each filter as the stacked photometry via,

\begin{equation}
    \bar{f_{i}}=\frac{\sum_{j}(w_{i,\,j}f_{i,\,j})}{\sum_{j}w_{i,\,j}},\,\,\bar{\sigma_i}=\sqrt{\sum_{j}1/w_{i,\,j}}
\end{equation}
where $w_{i,\,j}=1/\sigma_{i,\,j}^2$, $f_{i,\,j}$ and $\sigma_{i,\,j}$ denote the normalized flux and its associated uncertainty in the $i$ th filter of $j$ th galaxy\footnote{An alternative approach involves stacking the best-fit SEDs of individual galaxies. However, in certain sub-classified SOM cells, this method can yield unphysical composite SEDs, thereby compromising the reliability and generalizability of the resulting SED library. For this reason, we do not adopt this strategy.}. We emphasize that this weighted approach, though enhancing the S/N ratio, may potentially bias the stacked SEDs toward brighter and thus more massive galaxies. In Fig.~\ref{fig:error}, we quantify the differences between the mode stellar mass and SFR of individual galaxies in each SOM cell and the corresponding values derived from the stacked SED templates, denoted as $\Delta_{\rm mode-stack}$. We find only a slight offset typically below 0.1 dex, indicating that this effect has a negligible impact on our results.

Following the setup described in Sect.~\ref{subsec:jwst}, we perform SED fitting with BAGPIPES, ultimately synthesizing a total of  $30 \times 30$  SEDs. Then we perform photometry on the SEDs in the SED Lib using selective filters, constructing an equivalent new SOM that aligns with the test data’s filter set. In this test, we select 12 bands (e.g., $u,\ g,\ r,\ i,\ z,\ y,\ Y,\ J,\ H$, $K_{\rm s}$, ch1, ch2) from the COSMOS2020 catalog, consistent with those used in \citet{Davidzon_2022,Latorre_2024}, for demonstration and direct comparison. The new component map and color input are normalized by the standard deviation of the colors corresponding to the 900 SEDs in the SED Lib (including two empty cells, see Fig.~\ref{fig:new_som}).

As previously mentioned, when labeling physical quantities such as $M_\star$ and SFR -- both of which are influenced by the amplitude of the spectrum -- these quantities are normalized to F444W = 26. Then during parameter estimation, they are denormalized using the observed F444W magnitude. Since practical applications may lack F444W detections, the SED Lib is employed to correct the label maps via:

\begin{equation}
\label{eq:corr}
\begin{split}
    &X_{[x,y]} = \mathrm nX_{[x,y]} \times 10^{+0.4\times(26-\rm F444W_{[x,y]})} ,\\
    &\mathrm nX^*_{[x,y]} = X_{[x,y]} \times 10^{-0.4\times(m_0-m_{[x,y]})} 
\end{split} 
\end{equation}
where, in the pixel [$x,y$], ${\rm nX}_{[x,y]}$ and ${\rm nX}^*_{[x,y]}$ represent the normalized parameters in the original and corrected label maps, respectively. ${\rm F444W}_{[x,y]}$ and $m_{[x,y]}$ correspond to the photometric magnitude of the SED assigned to the given pixel, while $m_0$ is the reference magnitude. We tested different filter choices for normalization and found that using the band closest to F444W yields the most accurate estimates, likely because it minimizes errors introduced by the fitted SED during the label map correction process. Accordingly, we adopt the ch2 band as the normalization reference, with a fixed magnitude of $m_0 = 26$.

\subsubsection{Parameter estimation from the SED Lib}\label{subsec:est}

Now we have an adaptive SOM that galaxies can be directly mapped onto, then we move on to the mapping method. Following \citet{Latorre_2024}, we consider the impact of photometric uncertainties. Although in this work we find that the final statistic metrics of the parameter estimates are broadly consistent with those obtained using the five-point matching method in \citet{Davidzon_2019} and Sect.~\ref{subsec:test run}, this probabilistic framework allows new samples to be matched to the SOM in the form of a probability distribution, obtaining a representative average SED. For any given galaxy, the likelihood $\mathcal{L}$ of being mapped to a SOM pixel at coordinates [x,y] is defined via:
\begin{equation*}
    \mathcal{L}_{[x,y]} = e^{-\frac{\chi^2_{[x,y]}}{2}}
\end{equation*}
with:
\begin{equation}
    \chi^2_{[x,y]} = \sum_{N^*_{dim}} (\frac{(c_i-w^*_i)^2}{\sqrt{\sigma^2_{c_i}+\sigma^2_{\rm sys}}})_{\,[x,y]}\,\,,
    \label{eq:pdf}
\end{equation}
where $w^*_i$ is the $i$ th component of the updated SOM weights after SED replacement, $\sigma_{c_i}$ represents the photometric uncertainties of $c_i$, and $\sigma_{\rm sys}$ represent the parameter systematic uncertainty, which in this context is defined as $(\delta_{\rm M}/{\rm \log nM_{\rm label}})_{\,[x,y]}$. The likelihood surface is normalized by dividing $\mathcal{L}_{[x,y]}$ by its sum.
From these likelihood surfaces, combined with the pixel labels for a given parameter, we can derive probability distribution functions for various physical parameters.

   \begin{figure*}
        \centering
        \includegraphics[width=0.9\linewidth]{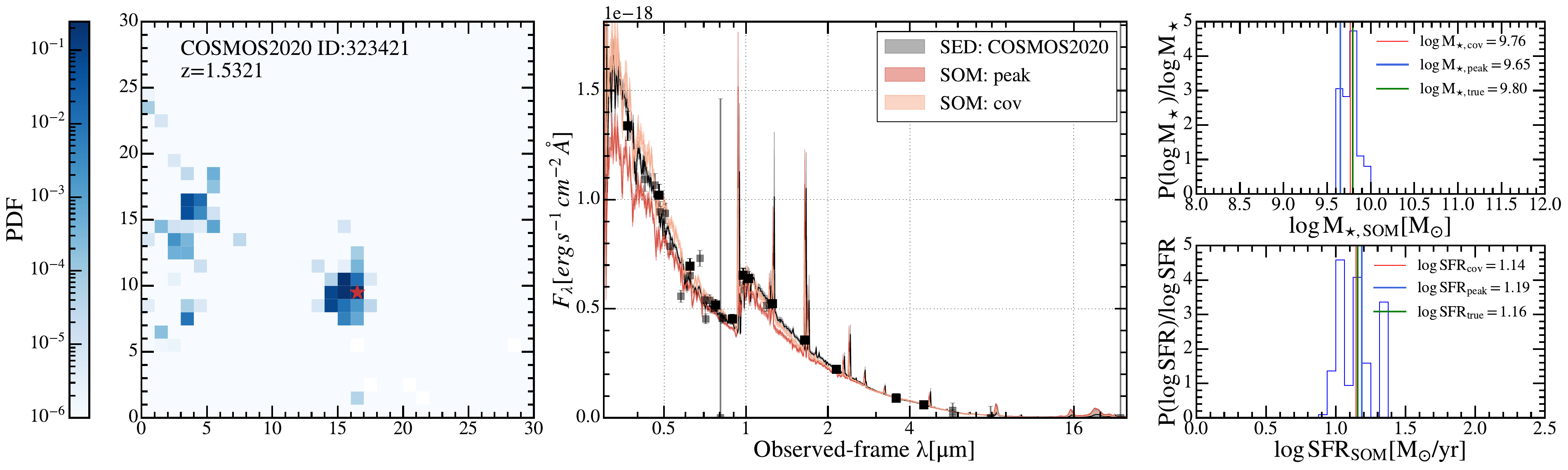}
        \caption{Example of parameter estimation using SHAPE. \textit{Left:} The probability distribution of galaxy ID323421. The target is mapped onto this surface according to the likelihood, with the red star marking the best-matching grid. \textit{Middle:} A comparison between the SEDs obtained via SED Lib matching (dark and light red) and the synthetic templates (black). The shaded regions represent the $16^{\rm th}-84^{\rm th}$ uncertainty range. Black dots with error bars denote the photometric points used as input ($u,\ g,\ r,\ i,\ z,\ y,\ Y,\ J,\ H,\ K_{\rm s}$, ch1, ch2), while transparent dots represent additional bands used in SED fitting but not included in SOM. \textit{Right:} The probability distribution of parameter estimates. The red vertical line represents the estimates obtained by convolving the probability distribution with the label maps, the blue line represents the estimate from the best-matching cell, and the green line corresponds to the SED fitting result.}
        \label{fig:pdf}
    \end{figure*}

\begin{table*}
    \centering
    \caption{Comparisons of parameter estimates for different catalogs using various methodologies. We report three key statistical metrics for the derived parameters, in the order of $M_\star$ and SFR. For fair comparison, the outlier fraction $\eta^*$ is defined as the fraction of objects with $|\log {\rm X_{SOM}}-\log {\rm X_{SED}}| > 0.5\,\rm dex$. The computational efficiency ($t_{\rm proc}$) of each model inevitably depends on factors such as CPU architecture and the complexity of the SED fitting templates, thus, we provide only an order-of-magnitude estimate as a reference. }
    \addtolength{\tabcolsep}{-0.2pt}
    \def\arraystretch{1.2}
    \begin{tabular}{lcccccc} 
	\hline
	Catalog & Band & Method & bias & $\sigma_{\rm NMAD}$ & $\eta^*\,[\%]$ & $t_{\rm proc}\,[\rm h/CPU/1000]$ \\
		\hline
            \textit{JWST}/PRIMER & F090W, ... F770W (9) & SOM25 & -0.03 | -0.04 & 0.12 | 0.18 & 0.7 | 7.3 & $\sim0.001$\\
            \textit{JWST}/PRIMER & F090W, ... F770W (9) & SHAPE25 & -0.03 | -0.05 & 0.16 | 0.18 & 2.8 | 7.3 & \\
            COSMOS2020/full-phot & $u,\ g,$ ... ch2 (12) & SHAPE30 & -0.03 | 0.01 & 0.19 | 0.25 & 3.1 | 11.6 & \\
            COSMOS2020/miss-phot & $u,\ g,\ \text{...}\ H$ (9) & SHAPE30 & -0.02 | -0.01 & 0.20 | 0.25 & 3.5 | 11.2 & \\
            COSMOS2020/miss-phot & $u,\ g,\ \text{...}\ H$ (9) & BAGPIPES & 0.19 | 0.18 & 0.22 | 0.35 & 19.2 | 14.1 & $\sim10$\\

		\hline
	\end{tabular}

    \label{table:results}
\end{table*}

Fig.~\ref{fig:pdf} presents an example of parameter estimation using SHAPE. In the left panel of Fig.~\ref{fig:pdf}, we show the probability distribution for a given galaxy. The middle panel presents a comparison of SEDs obtained using two different matching methods with 11 colors as input, against the SED derived from full-band SED fitting. The dark red solid line represents the SED from the SED library that best matches the input colors, while the light red solid line corresponds to the probability-weighted SED. The black solid line represents the SED fitting result. The shaded region encompasses the 1 $\sigma$ uncertainty range. Black dots with error bars represent the photometric points used as input, while transparent dots indicate additional photometric points used in the SED fitting but not included in SOM.
In the right panel, we present the probability distribution of parameter estimates. The red vertical line represents the estimates obtained by convolving the probability distribution with the label maps, the blue line represents the estimate from the best-matching cell, and the green line corresponds to the SED fitting result. The comparison demonstrates that the matched SEDs and estimated parameters closely align with the SED fitting results, further validating the effectiveness of our approach.

\section{Results} \label{sec:results}

\subsection{Stellar mass and SFR estimates}\label{subsec:results}
In this section, we present the results of parameter estimation using SHAPE. Direct comparisons are shown in Fig.~\ref{fig:results} and Table~\ref{table:results}. 

Before directly applying SHAPE to the COSMOS2020 test data, we first check whether it's still effective compared to the traditional SOM method by evaluating the estimation quality in the 7:3 sample split of the JWST catalog in Sect.~\ref{subsec:test run}. Strictly following the procedures outlined in Sect.~\ref{subsec:SHAPE}, we construct a $25\times25$ SED Lib and select the JWST bands (F090W, F115W, F150W, F200W, F277W, F356W, F444W, F410M and F770W) to reconstruct the component maps.
For batch processing, we perform parameter estimation by convolving the probability distribution with the label map. To mitigate potential biases, we exclude cells with probabilities below 1\%. 
In the middle panels of Fig.~\ref{fig:results}, we present a comparison of the stellar mass and SFR estimates obtained using the $25\times25$ $\rm SOM^*$ with those derived from SED fitting. For each estimate, four statistical metrics are evaluated. The boundaries of outliers are indicated by red dashed lines. Both estimates show good agreements with the SED fitting results, and only a very slight increase in $\sigma_{\rm NMAD}$ and bias is observed, indicating that the error introduced by the SED Lib is minimal.

In the lower panels of Fig.~\ref{fig:results}, we present the estimation results of applying $30\times30$ $\rm SOM^*$ in Sect.~\ref{subsec:sed} to the COSMOS2020 dataset through SHAPE.
In the left panel, the $M_\star$ estimates exhibit a strong agreement with the SED fitting results, despite the SOM method utilizing fewer photometric bands. The standard deviation of the differences is within 0.2 dex, with only 5.6\% of outliers. No significant bias is observed, which can be attributed to the completeness of the training sample based on the JWST PRIMER catalog. This ensures that even low-mass galaxies in the COSMOS2020 catalog can find well-matched SEDs within the SED Lib.

In the right panel, the SFR estimates show slightly larger deviations compared to the stellar mass estimates but still demonstrate a strong 1:1 correlation with the SED fitting results, with 0.25 of $\sigma_{\rm NMAD}$ and 5.4\% of outliers. A decline in accuracy is observed at the high-SFR end. This behavior arises due to the fact that, within SOM, the classification accuracy for specific galaxy types and the resolution of parameter estimation (i.e., the ability to distinguish between different parameter combinations) are influenced by the representation of such galaxies in the training set. Galaxies with lower representation correspond to less SOM cells, leading to decreased precision in parameter estimates. Specifically, the JWST catalog contains a larger proportion of low-$M_\star$ and low-SFR galaxies, enhancing the sensitivity of the model to such populations. However, for galaxies experiencing strong dust attenuation -- primarily located in the lower right region of the SOM -- their SEDs are affected by prominent emission lines in certain bands, introducing intrinsic complexity. A sufficiently large number of such galaxies is required to accurately account for dust emission effects. Another challenge arises from the incomplete coverage of galaxies in the FIR and sub-mm, as not all galaxies have detections in these wavelengths. Consequently, SFR estimates derived from SED fitting may exhibit inaccuracies in such cases.

\subsection{Comparison to synthetic templates}\label{sec:vs}

\begin{figure*}
        \centering
        \includegraphics[width=0.4\linewidth]{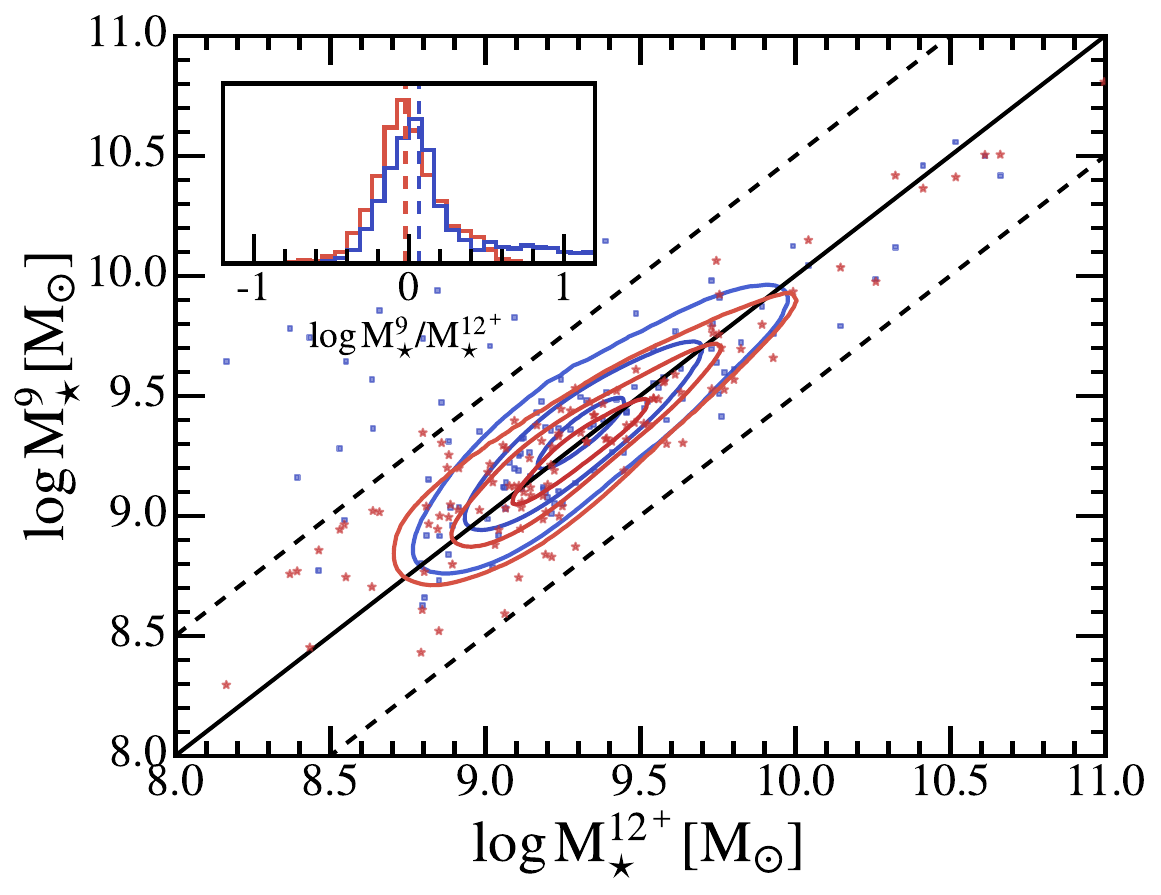}
        \includegraphics[width=0.4\linewidth]{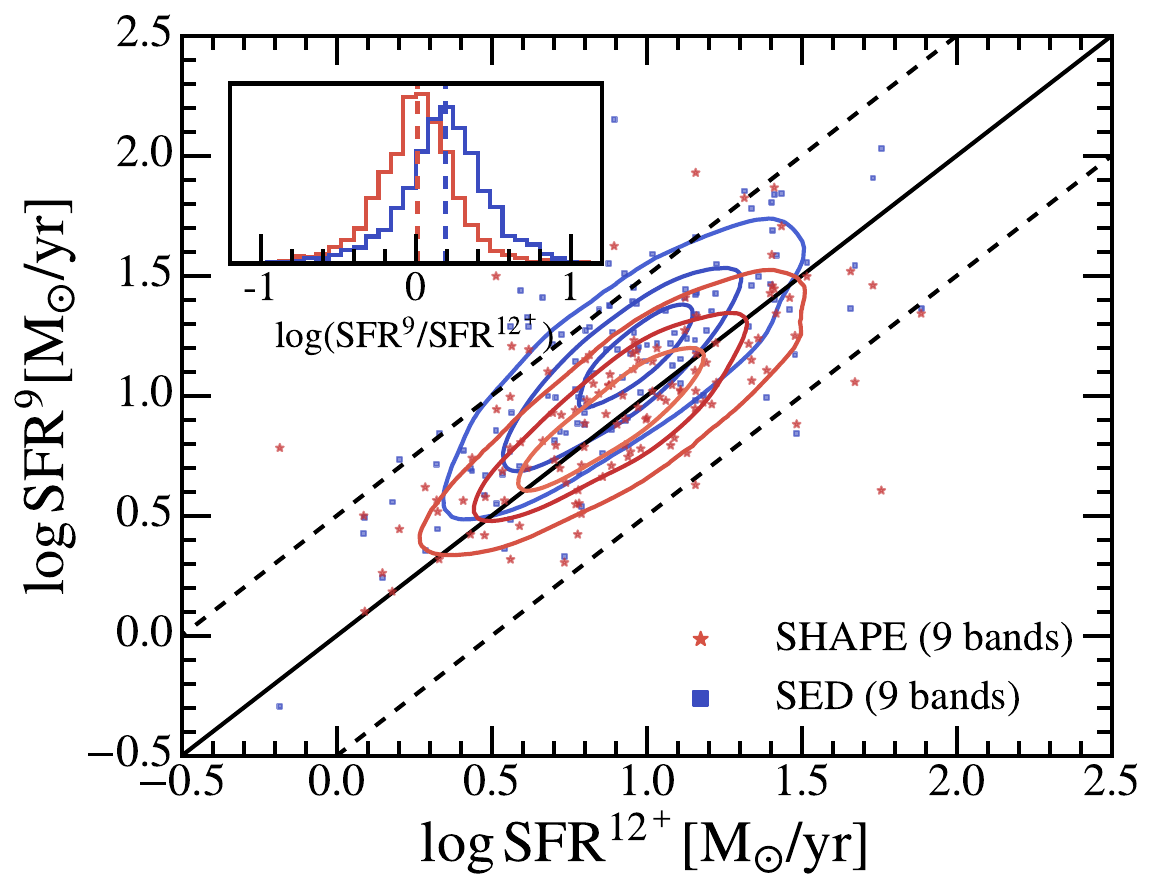}
        \caption{Comparison between SOM- and SED-derived estimates using the 9-band photometry (red and blue, respectively), with 12+-band photometry values as a reference. The solid line represents the zero offset, while the two dotted lines indicate ±0.5 dex boundaries. A random selection of 100 galaxies is used for visualization. The same color scheme is applied to the histograms in the inset, showing the respective distributions.}
        \label{fig:cosmos2}
\end{figure*}
\begin{figure*}
        \centering
        \includegraphics[width=0.9\linewidth]{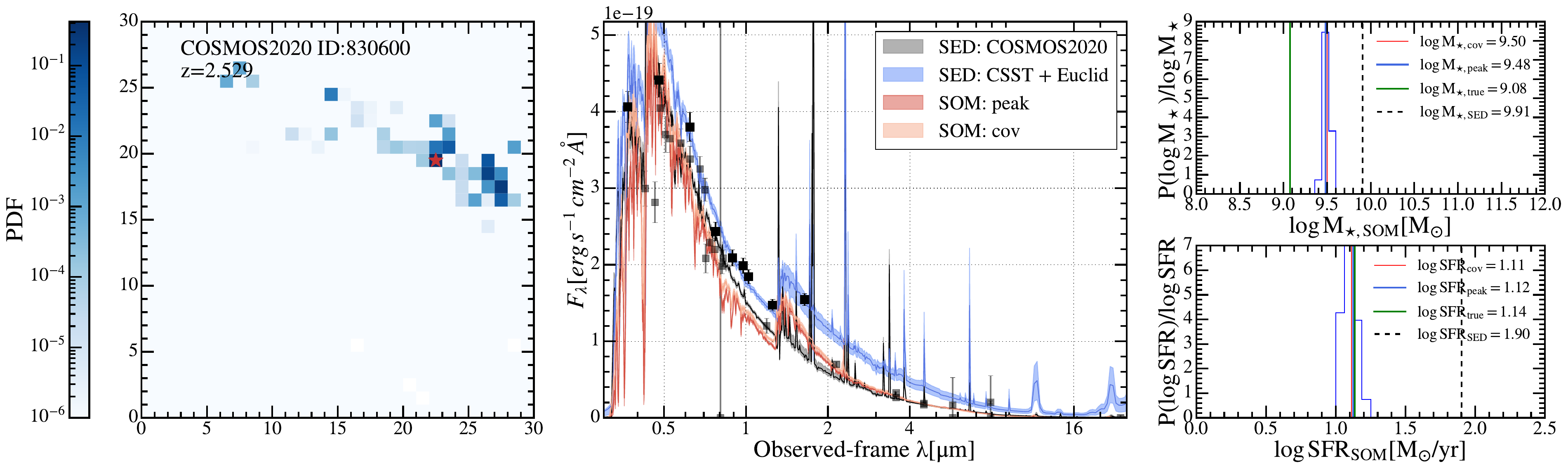}
        \caption{Similar to Fig.~\ref{fig:pdf}, but using only $u,\ g,\ r,\ i,\ z,\ y,\ Y,\ J,\ H$ bands in SOM. The blue line in the middle panel represents the synthetic templates obtained using the same 9-band photometry. The dashed lines in the right panel correspond to the 9-band SED fitting estimates.}
        \label{fig:pdf2}
\end{figure*}
To prepare for future large-scale surveys such as CSST, we assess the performance of SHAPE and SED fitting for parameter estimation using a fixed set of nine photometric bands that correspond to the observational bands of CSST and \textit{Euclid}. These estimates are then compared to those obtained from SED fitting using full photometric coverage. In this analysis, we select the nine bands (e.g., $u,\ g,\ r,\ i,\ z,\ y,\ Y,\ J,\ H$) and repeat the procedures outlined in Sect.~\ref{subsec:est} to obtain SOM-based estimates. Here we adopt the $H$ band for label map correction via Eq.~\ref{eq:corr}, as the ch2 band is not available in CSST and \textit{Euclid} photometric systems. For a fair comparison, we also perform SED fitting with the same nine bands with no ancillary data to derive a second set of SED-based estimates (see Sect.~\ref{subsec:csst}). In this analysis, FIR and sub-mm detections are not explicitly required, as such selection criteria would introduce a bias toward high-SFR and heavily dust-attenuated galaxies, which are underrepresented in the JWST PRIMER field. However, by incorporating additional bands such as $K_{\rm }s$, ch1, ch2, and medium/narrow bands, SFR estimates derived from 12+ bands can be considered more robust compared to those obtained using nine-band photometry alone.

In Fig.~\ref{fig:cosmos2}, we compare the SED- and SHAPE-derived estimates using 12+-band SED fitting results as a reference. The red contours represent the distribution of SHAPE-based estimates, while the blue contours denote those obtained from SED fitting. To avoid excessive overlap, we randomly select 100 galaxies for visualization. For each estimate, we calculate the discrepancies between the estimated values and the true values, present them in the inserted panel. In the left panel, both SED- and SHAPE-derived $M_\star$ show no significant offset from the reference values, although SED fitting displays a subset of galaxies with overestimated $M_\star$, a trend not observed in the SOM-based estimates. In the right panel, $\rm SFR_{SOM}$ aligns well with the reference values at the low-mass end with a bias of -0.01, whereas SED fitting shows a systematic bias of 0.18 dex due to the exclusion of the $K_{\rm s}$, ch1, and ch2 bands. For both parameters, the scatter of SOM-based estimates is slightly lower than that of the SED-based estimates.

Specifically, in Fig.~\ref{fig:pdf2}, we present an example comparing the SED obtained through SOM matching with the SED derived from SED fitting. Since our SOM is imprinted with MIRI band information during clustering and labeling, the SOM-derived SED (red lines) demonstrates better consistency with the reference SED (black lines) at longer wavelengths compared to the SED derived from template fitting (blue lines). Quantitatively, when using the exact same filter set, 60\% of galaxies have SOM-derived SFR values closer to the reference than those obtained from SED fitting. These results confirm the effectiveness of the SOM-based method for future large-scale survey projects.

\section{Discussions} \label{sec:discussions}
\subsection{Comparisons to previous work}

This study builds upon the foundational work of \citet{Davidzon_2019, Davidzon_2022}, with several key extensions. A major distinction from previous studies lies in our focus on utilizing JWST as a benchmark and bridging data-driven techniques with traditional SED fitting, integrating observational data with template-based modeling.

Despite employing a smaller SOM size -- corresponding to a lower parameter resolution -- and incorporating SED fitting (which introduces certain model-fitting uncertainties), our model performs comparably as the approach
presented in \citet{Davidzon_2022}. The improvement is attributed to JWST’s superior imaging quality and the inclusion of MIRI data. Specifically, for the $M_\star$ estimates, \citet{Davidzon_2022} reported a  symmetric scatter of 0.25 dex and that approximately 3\% of galaxies exhibited significant underestimation (by more than a factor of 3), whereas in our work, the scatter is 0.20 dex and no significant bias or extreme outliers are observed. We cannot directly compare the quality of SFR estimates, as the labeling methods differ significantly.

For parameter estimation, we adopt the probabilistic distribution approach proposed by \citet{Latorre_2024}, effectively addressing the issue of missing data at this stage. However, during the model training phase, we do not follow their method of randomly sampling fluxes. This study primarily focuses on validating the methodology, and in future work, we plan to incorporate their approach to better address missing data during training, thereby expanding both the training sample and the SOM size.

In our study, the introduction of the SED Lib appears to act as an additional source of uncertainty. This is largely due to the limited sample size, which results in many SOM grid cells representing mixed or under-resolved subclasses of galaxies, i.e., different types of galaxies are not fully separated into distinct cells, and some types are too sparsely populated to form independent, well-defined nodes. When constructing the SED Lib, stacking SEDs within these under-resolved cells makes it difficult to capture the intrinsic features of each subclass. However, as the dataset expands and galaxy classification becomes more refined, SED Lib may transition from being a source of error to serving as a physical constraint, mitigating the impact of photometric uncertainties and potentially improving the overall accuracy of parameter estimation.

\subsection{Limitations of this work}\label{subsec:limit}
The primary limitation of this study arises from the restricted size of the training sample. Both \citet{Davidzon_2022} and \citet{Latorre_2024} have demonstrated that an 80×80 SOM provides an optimal compromise. Given the higher redshift accuracy and broader stellar mass ($M_\star$) range of the JWST sample, a larger SOM size would ideally be required. However, due to our stringent selection criteria, the number of sources in our training set is significantly smaller than in previous studies (7,507 galaxies compared to 174,522 and 228,524, respectively). We adopt a relatively small 30×30 SOM in the initial clustering step. This choice sacrifices classification accuracy and reduces the resolution of parameter estimation. The impact is most pronounced at the high-SFR end, where the number of massive dusty star-forming or starburst galaxies is only sufficient to populate approximately 50 grids. Consequently, dust emission is not well resolved in the current model. However, we anticipate that the model will be continuously refined as more data become available.

Another limitation associated with the relatively small data sample is that, although JWST reaches much deeper imaging depths and is theoretically capable of detecting smaller and fainter galaxies, the COSMOS2020 survey covers a substantially larger field. As a result, the latter includes a significantly higher number of very massive or extreme galaxies compared to the JWST training set, implying that the training sample may not be fully representative. Nevertheless, this issue is secondary, as SOMs are inherently limited in accurately estimating the properties of such extreme objects; these galaxies typically fall into peripheral cells with suboptimal clustering performance—commonly referred to as the boundary effect \citep{Davidzon_2019}.

In addition, challenges may arise form the treatment of missing data and upper limits during the training process. This lack of a dedicated solution restricts the size of the training sample and potentially biases the selection, which affects the representativeness of the trained SOM. Specifically, requiring a non-zero detection in F090W may inadvertently exclude heavily dust-obscured galaxies, while imposing a high S/N requirement for F770W may result in the exclusion of low-mass galaxies. However, as we constrain our sample to the range $1.5 < z < 2.5$, the selection effects should be relatively limited.
Moreover, addressing missing data ultimately involves the prediction of absent fluxes, and clustering itself can be utilized as a potential method for flux prediction. In our forthcoming SHAPE II framework (Wang in prep.), we're developing an iterative approach wherein parameter estimation and flux prediction are refined cyclically through clustering, progressively enhancing the accuracy and completeness of the model.

\subsection{Redshift estimation through SOM}
Throughout this work, we evaluate the capability of SOM and SHAPE to estimate stellar mass and SFR, while noting that both methods underperform in recovering photometric redshift. Although some simulation-based studies~\citep{Davidzon_2019,Latorre_2024} have argued that SOMs can yield reliable redshift estimates, empirical applications trained on observational data tend to face similar challenges. For example, \citet{Davidzon_2022} did not show the redshift estimates but fixed the redshifts of their sources prior to estimating other physical properties. \citet{Abedini_2025} also reported that recovering photometric redshifts using SOMs remains challenging.

Our strategy is also to treat photometric redshift as a prior and use it to bin the data, rather than as a target parameter for direct inference. This approach is motivated by two key considerations. First, SOMs are primarily sensitive to the continuum shape of galaxy SEDs and identify redshifts through broad spectral features such as the Lyman and Balmer breaks; they are generally less effective than SED fitting at capturing narrow emission-line information\footnote{\citet{Masters_2015} and \citet{Zhang_2025} successfully employed SOMs to calibrate photometric redshifts, but adopted a distinct methodology, and is therefore beyond the scope of this study.}. Second, reliable redshift estimation via SOM requires sufficient sampling density in the 3D $z$–$M_\star$–SFR space. That is, a SOM node must be populated by at least two galaxies with nearly identical SEDs and redshifts to define a meaningful cluster, a requirement not met by the current JWST training sample. An alternative approach
would be to incorporate spectroscopic redshifts to refine photometric redshift estimates~\citep[e.g.][]{Masters_2015,Hemmati_2019,Zhang_2025}.

\section{Conclusions} \label{sec:conclusions}
In this study, we explore the application of SOM to bridge different photometric systems, thereby leveraging the exceptional depth, quality, and wavelength coverage of the JWST PRIMER survey to calibrate the estimation of physical parameters in upcoming wide-field surveys. Below, we summarize our main findings:

\begin{enumerate}
    \item The SOM trained on JWST data provides high accuracy in parameter estimation. The SOM achieves parameter estimates that are nearly identical to those obtained from SED fitting, with $\sigma_{\rm NMAD}<0.2$ for both $M_\star$ and SFR, while offering significantly higher computational efficiency.
    
    \item We introduce a hybrid method called SHAPE which replaces the default SOM weights with SED Lib, derived from galaxies in each cell. Utilizing the SED Lib, the SOM model can be extended to samples from different filter sets, such as COSMOS2020 and CSST. This enables parameter estimation even when the input color combinations do not exactly match the training data, as the SED Lib provides a continuous framework to replace discrete photometric points. This advancement significantly enhances the generalizability of the SOM method across multiple surveys, marking a major step forward in multi-survey parameter estimation.
    
    \item We apply this method to JWST and COSMOS2020 test sample. For the JWST catalog, $\sigma_{\rm NMAD}$ increase slightly to 0.16 for stellar mass estimates. Therefore, the error introduced by SED Lib is proved minimal. For the COSMOS catalog, it demonstrates a strong agreement between $M_{\star,\ \rm SOM}$ and $M_{\star,\ \rm SED}$. While SFR estimates exhibit slightly larger discrepancies at the high end, the overall deviations remain within 0.5 dex, validating the reliability of our approach.  

    \item The SOM suggests its promising application in upcoming large-scale surveys, with its outstanding robustness and efficiency. With a limited set of bands (e.g. $u,\ g,\ r,\ i,\ z,\ y,\ Y,\ J,\ H$), our method produces unbiased SFR estimates than traditional SED fitting (with a bias of -0.01 and 0.18 respectively).
    
\end{enumerate}

Overall, this work, though based on a relatively small JWST sample, already demonstrates the potential of SHAPE in efficiently processing and analyzing large, high-dimensional astronomical datasets. In our next-generation of SHAPE II, future efforts will focus on addressing challenges from missing data and extending the model to other redshift bins, particularly through the incorporation of more comprehensive datasets as they become available from JWST (e.g. COSMOS-Web; \citealt{Casey_2023,Shuntov_2025}). With a hopefully more complete and robust SHAPE II, we aim to explore its utility in guiding future observations, identifying peculiar objects, and predicting fluxes across multiple bands.

\begin{acknowledgements}
      This work was supported by National Natural Science Foundation of China (Project No.12173017 and Key Project No.12141301), National Key R\&D Program of China (grant no. 2023YFA1605600), Scientific Research Innovation Capability Support Project for Young Faculty (Project No. ZYGXQNJSKYCXNLZCXM-P3),  and the China Manned Space Program with grant no. CMS-CSST-2025-A04.
\end{acknowledgements}

\bibliographystyle{aa}
\bibliography{ref}

\appendix
\setcounter{figure}{0}
\renewcommand{\thefigure}
{A\arabic{figure}}

    \begin{figure*}
        \centering
        \includegraphics[width=0.22\linewidth]{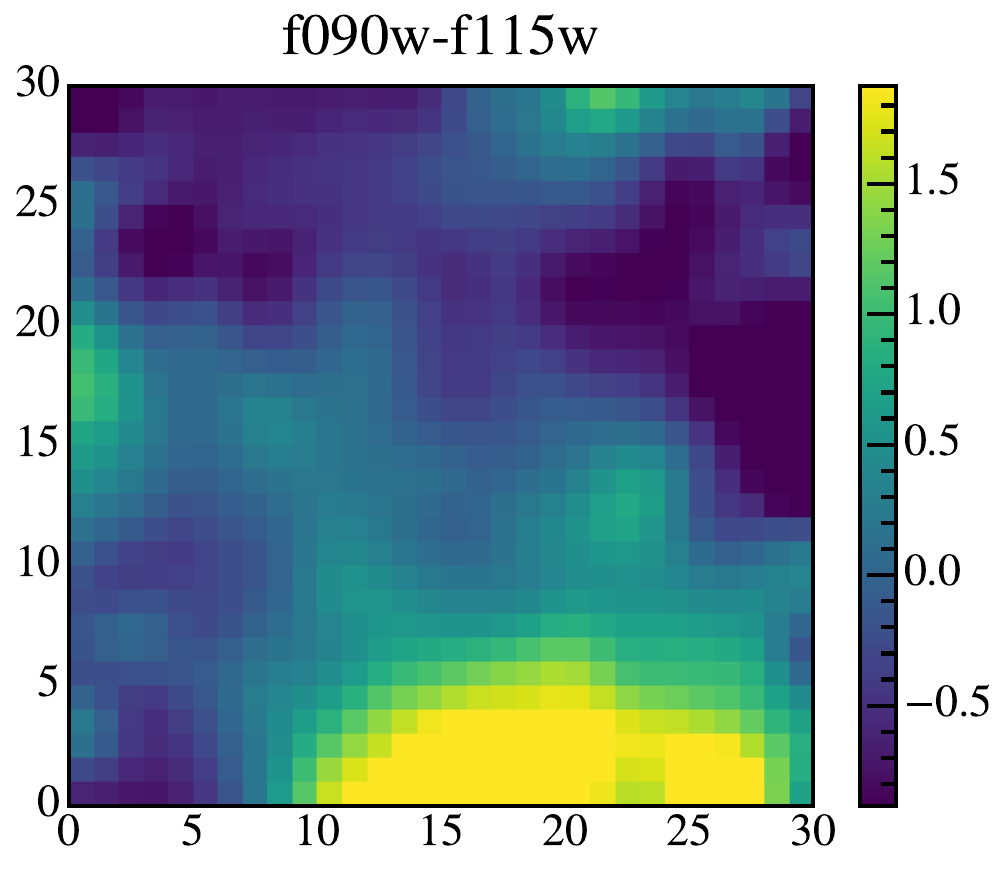}
        \includegraphics[width=0.22\linewidth]{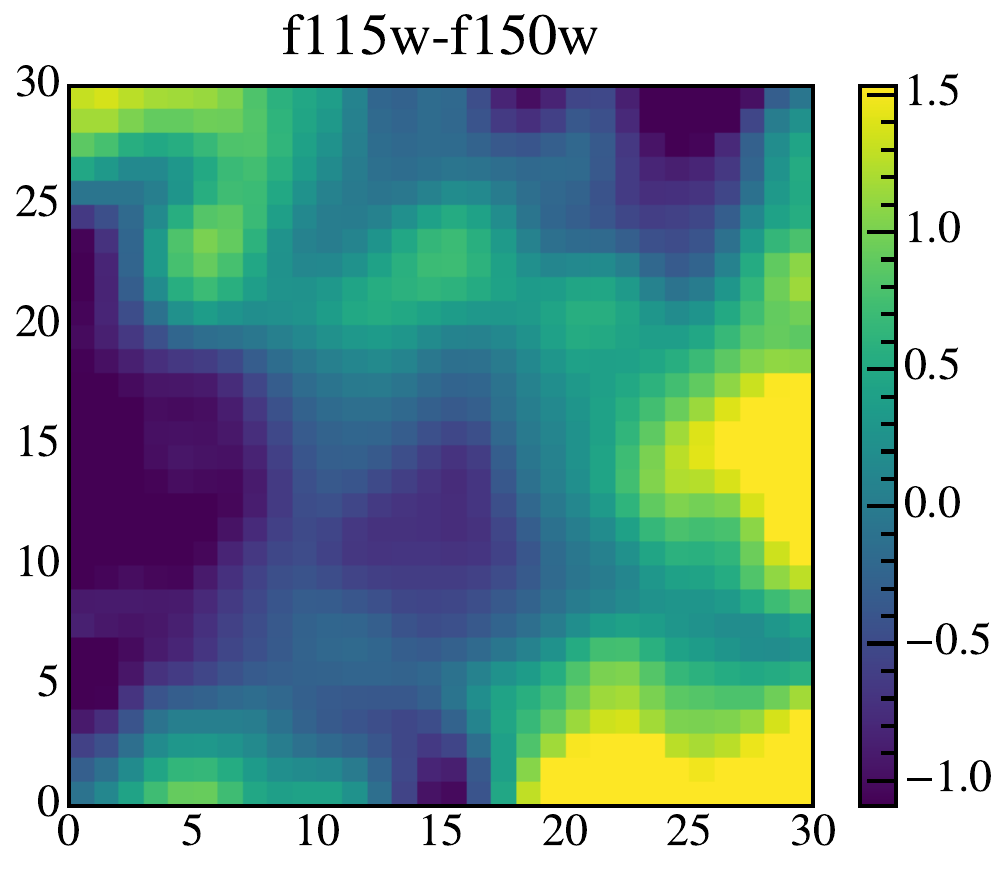}
        \includegraphics[width=0.22\linewidth]{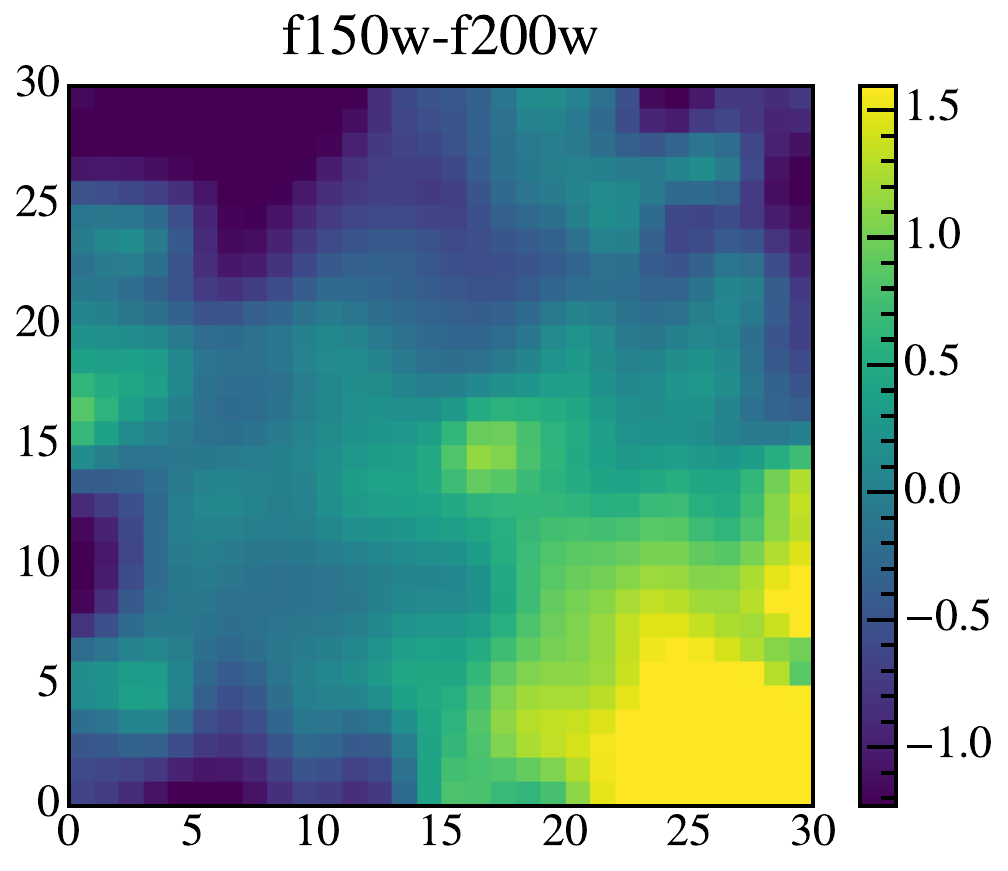}
        \includegraphics[width=0.22\linewidth]{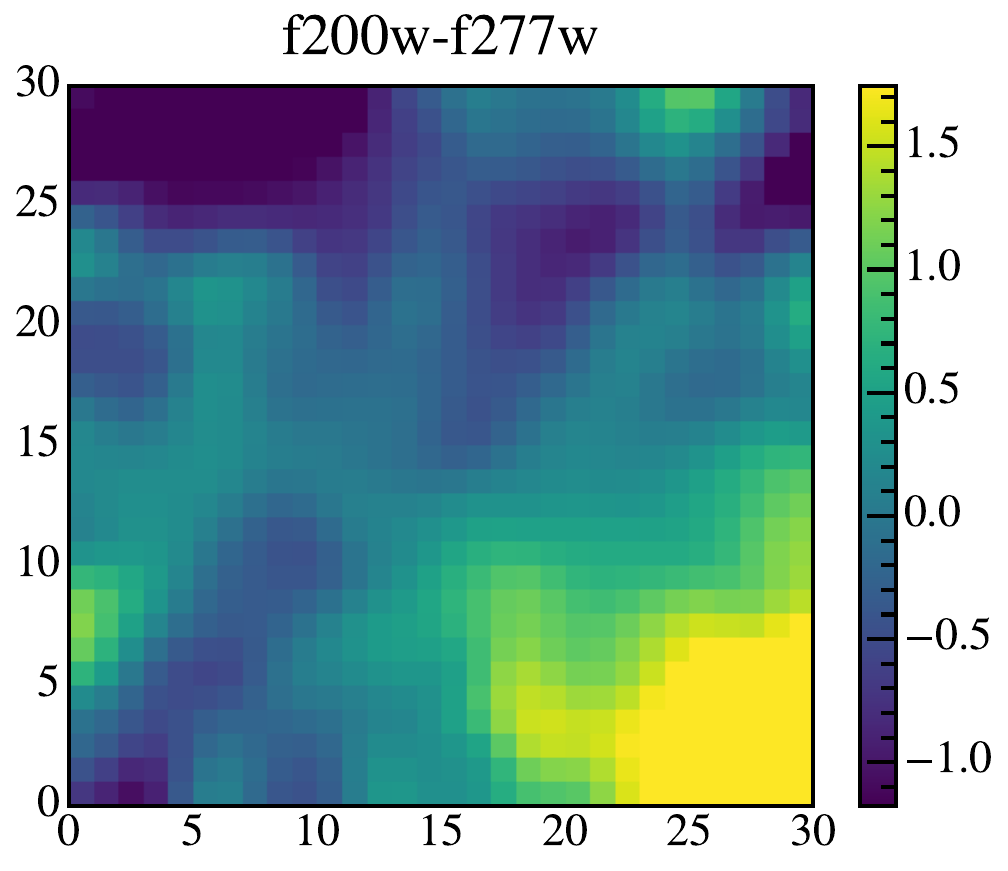} \\
        \includegraphics[width=0.22\linewidth]{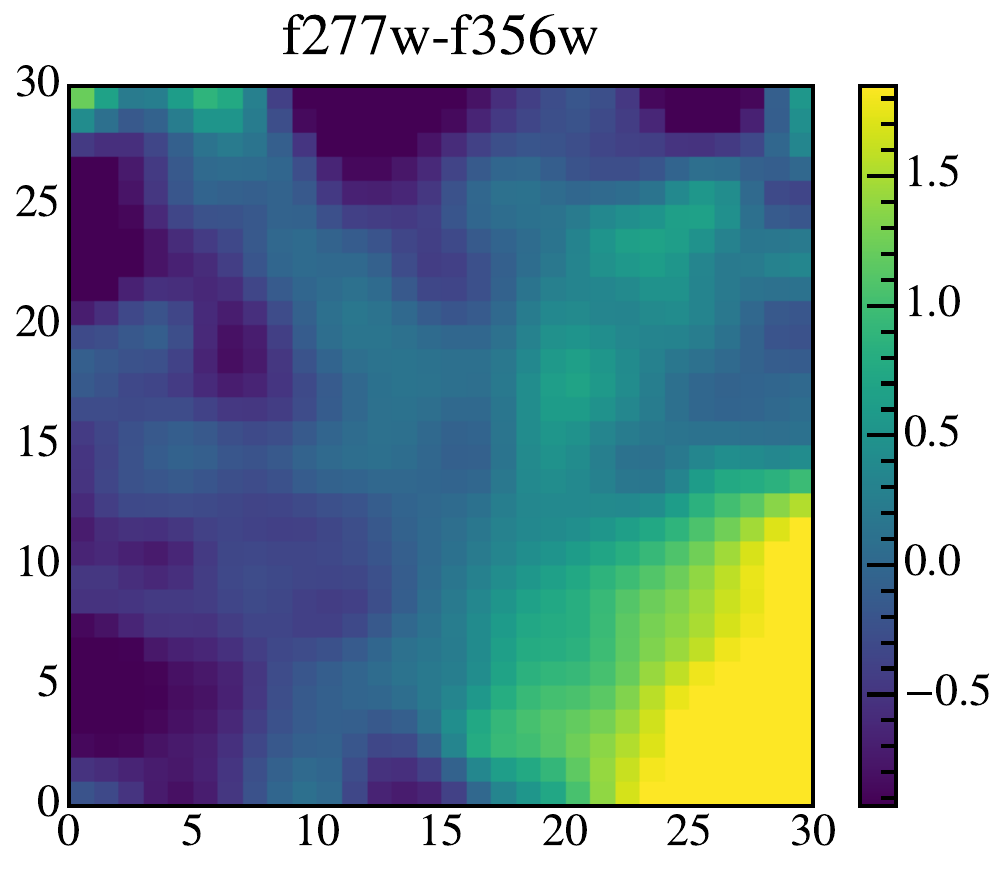}
        \includegraphics[width=0.22\linewidth]{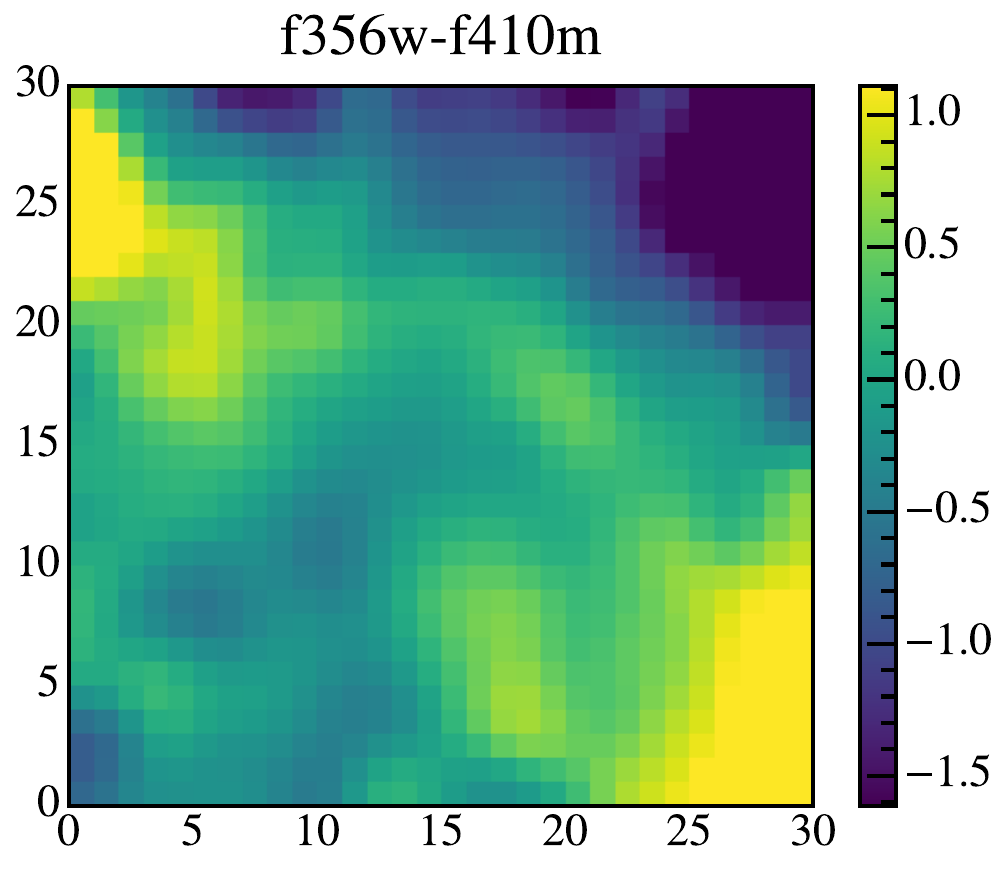}
        \includegraphics[width=0.22\linewidth]{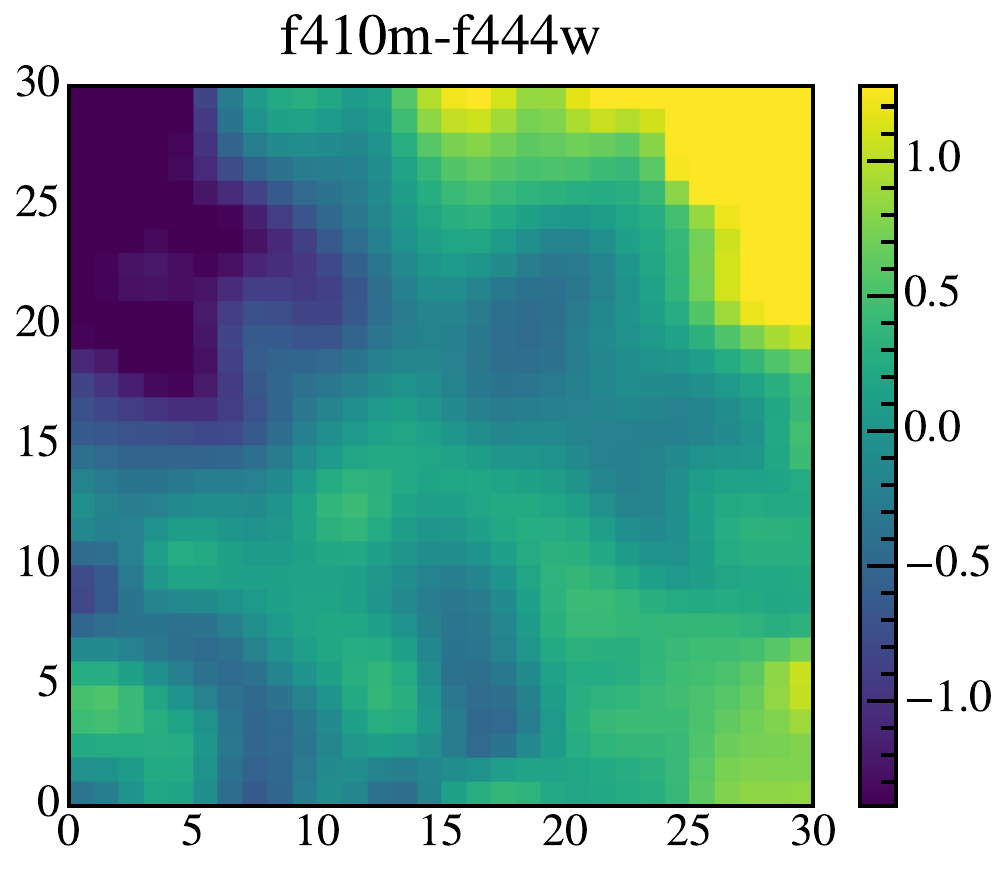}
        \includegraphics[width=0.22\linewidth]{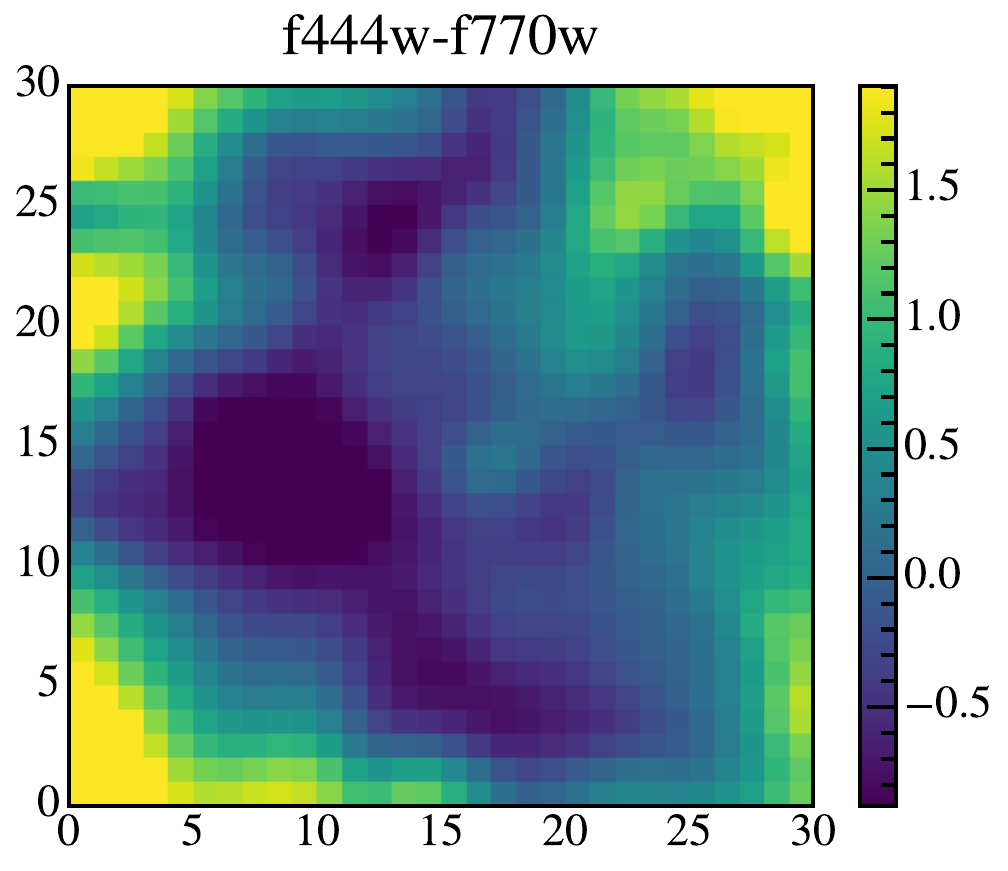}
        \caption{Component maps showing each of the 8 input colors across our trained SOM. The color bar indicates the normalized color values.}
        \label{fig:colors}
    \end{figure*}

    \begin{figure*}
        \centering
        \includegraphics[width=0.22\linewidth]{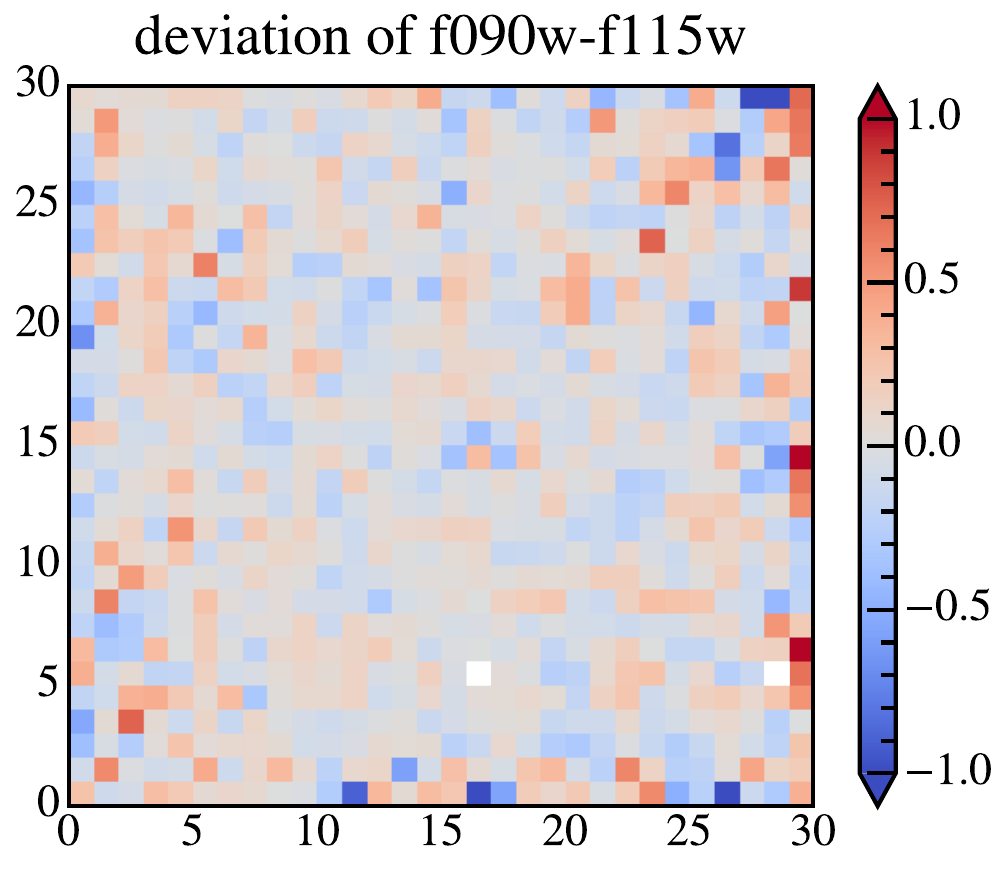}
        \includegraphics[width=0.22\linewidth]{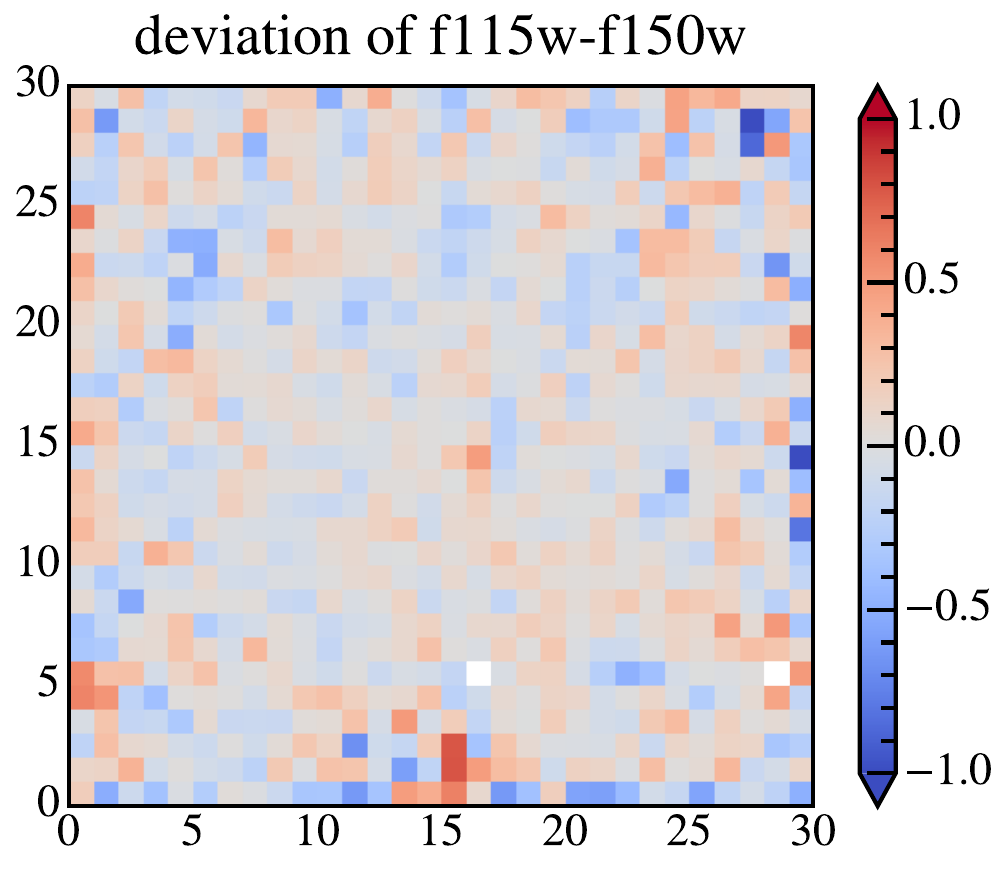}
        \includegraphics[width=0.22\linewidth]{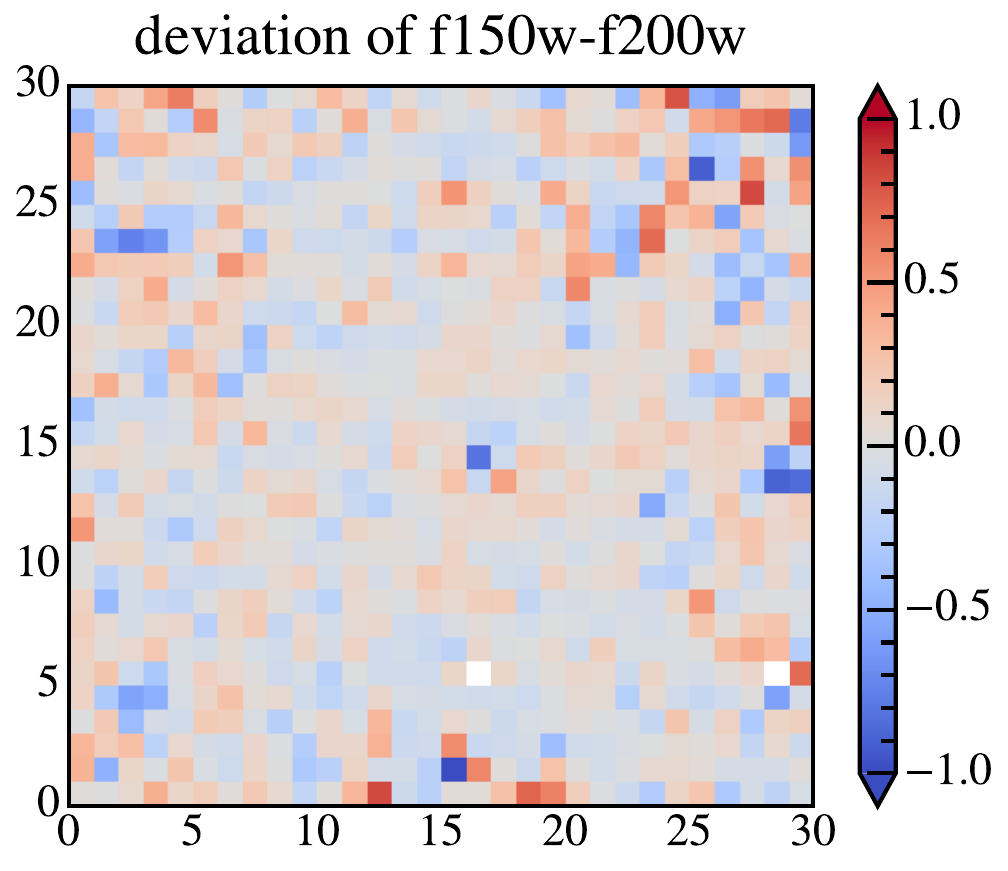}
        \includegraphics[width=0.22\linewidth]{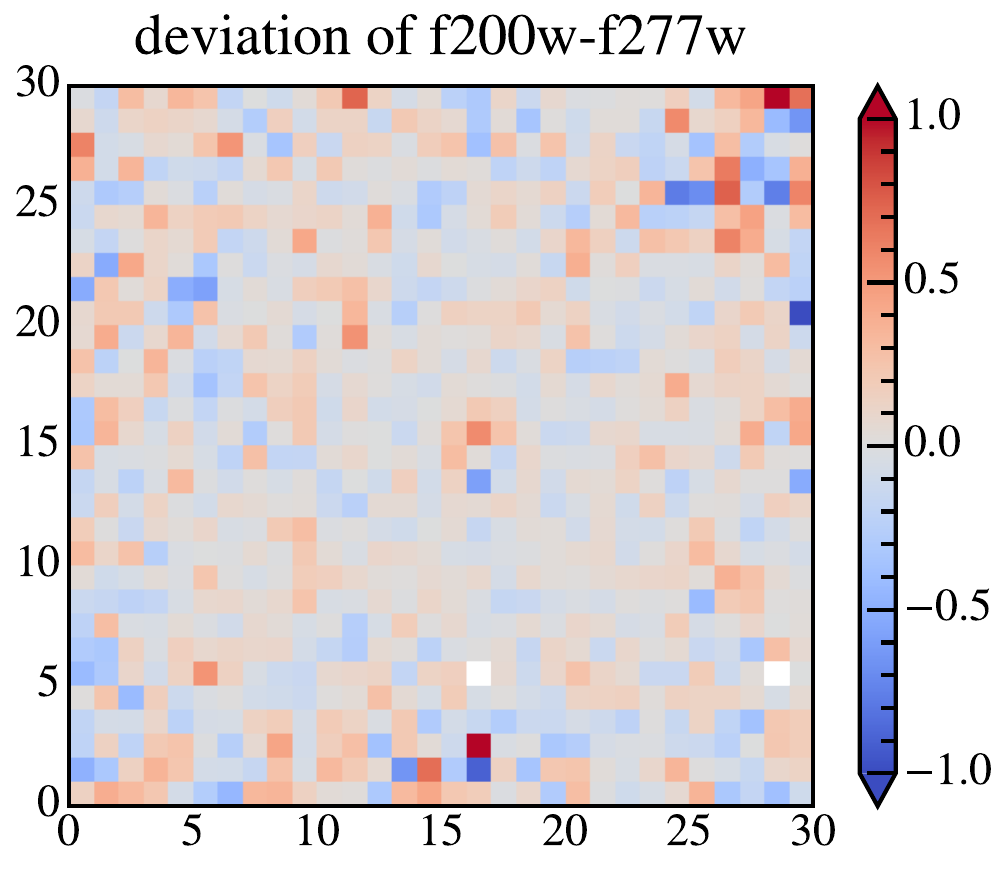} \\
        \includegraphics[width=0.22\linewidth]{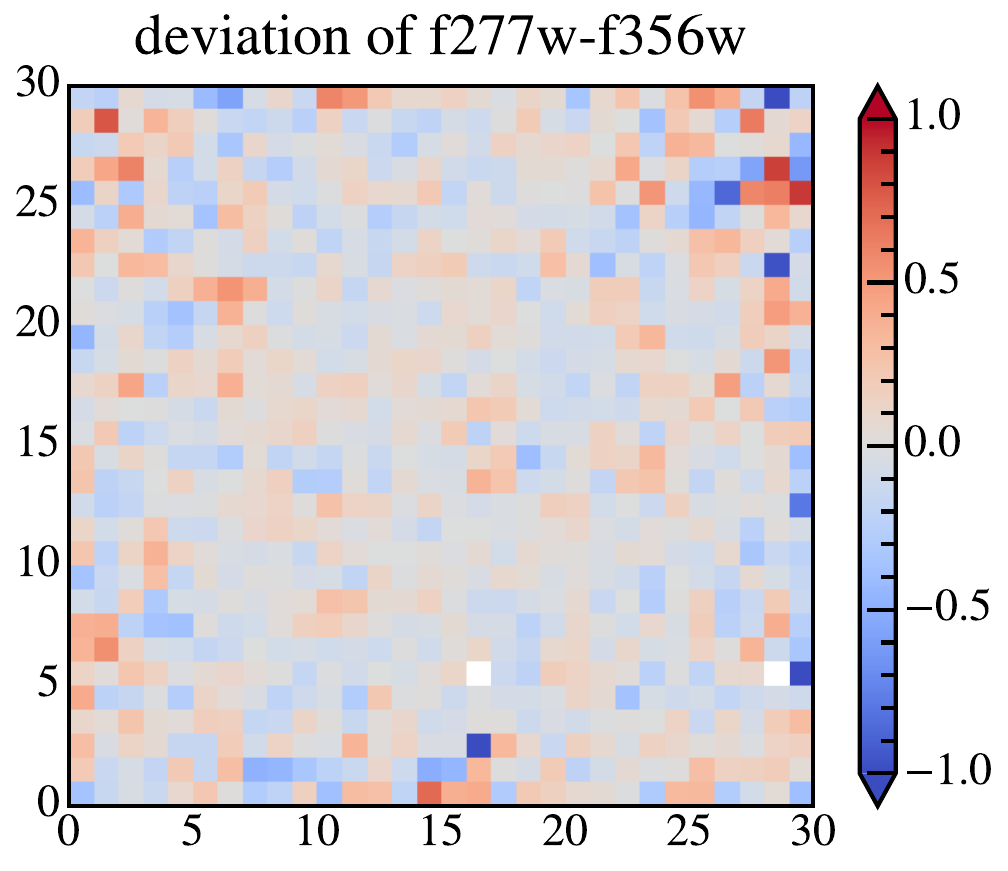}
        \includegraphics[width=0.22\linewidth]{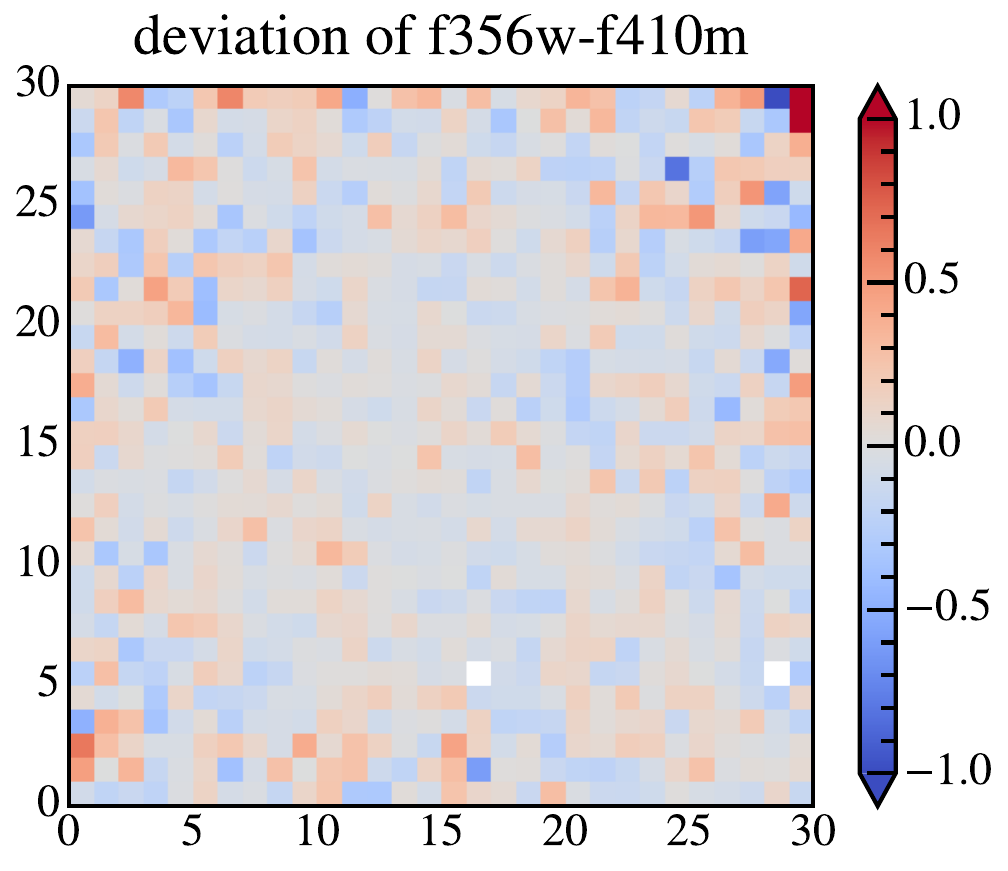}
        \includegraphics[width=0.22\linewidth]{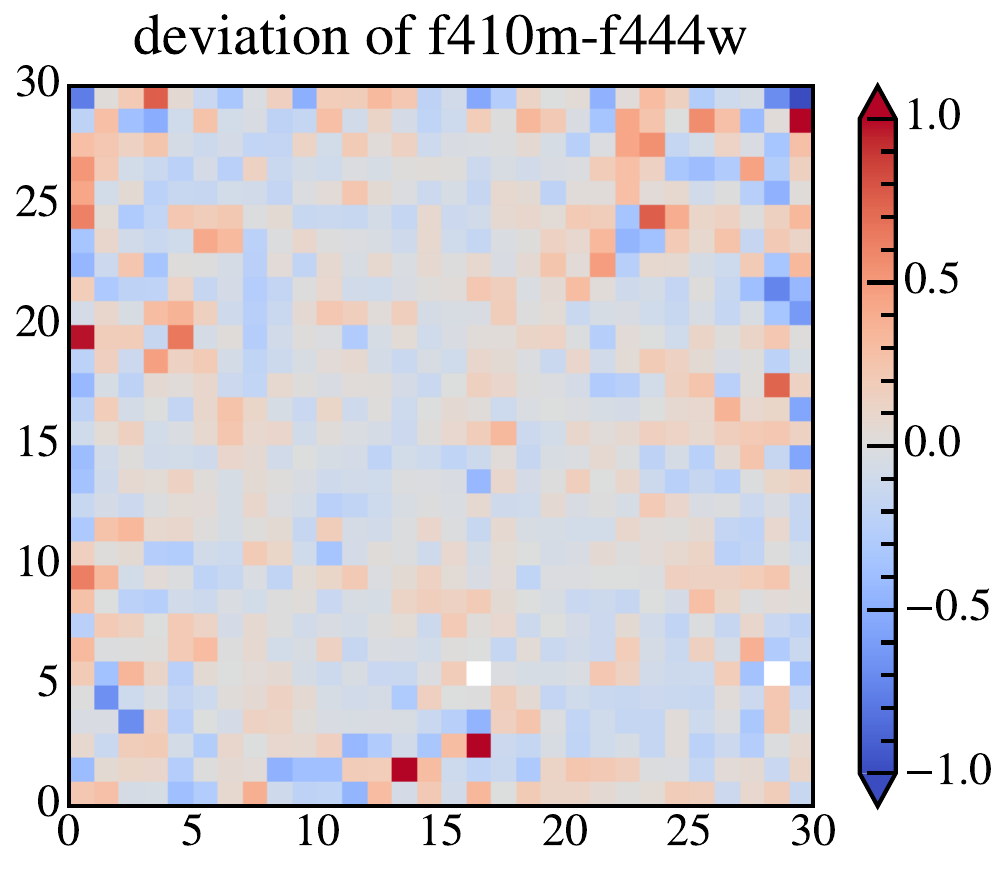}
        \includegraphics[width=0.22\linewidth]{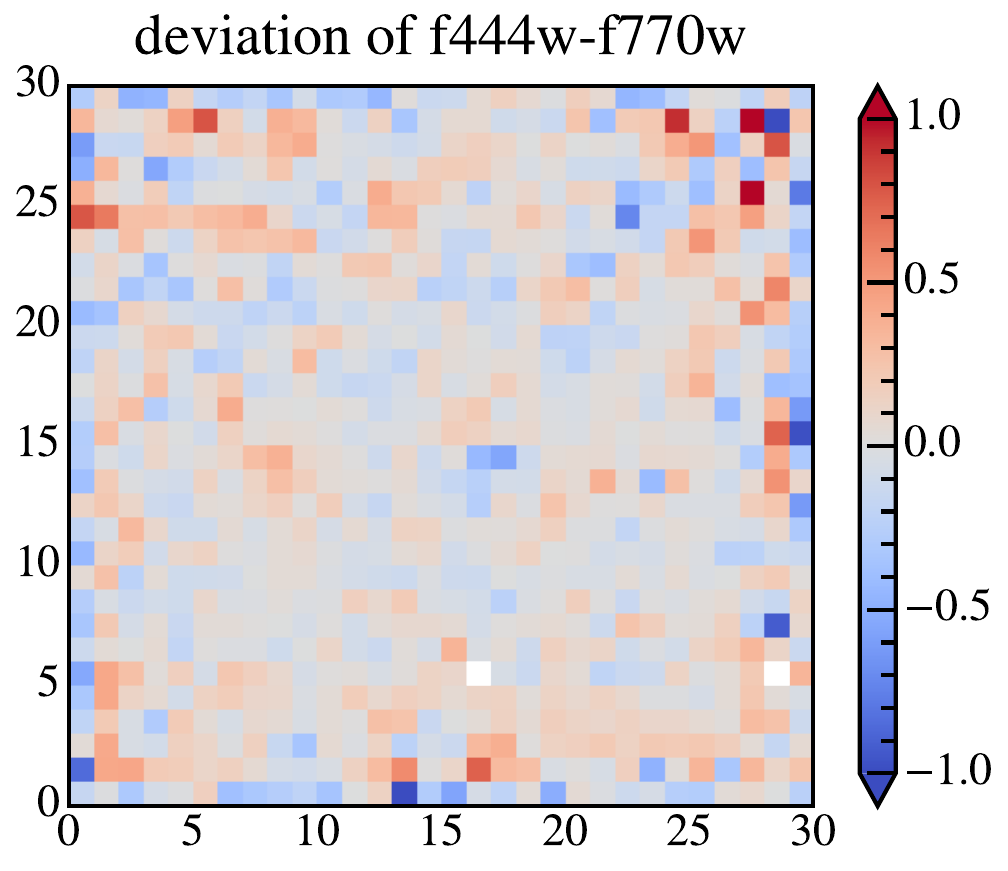}
        \caption{Deviation of SOM-generated weights and median normalized colors of galaxies of each.}
        \label{fig:dev}
    \end{figure*}

\begin{figure*}
    \centering
    \includegraphics[width=0.35\linewidth]{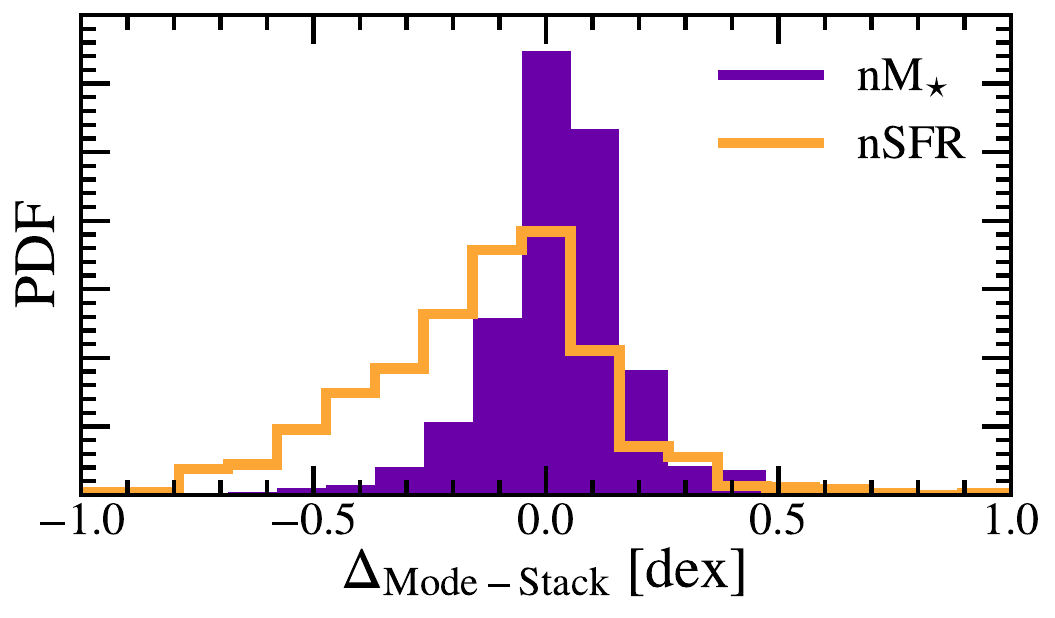}
    \caption{Distribution of $\Delta_{\rm mode-stack}$, defined as the difference between the mode $M_{\star}$ and SFR of individual galaxies in each SOM cell and the corresponding values derived from the stacked SED template.}
    \label{fig:error}
\end{figure*}
    
\begin{figure*}
    \centering
    \includegraphics[width=0.15\linewidth]{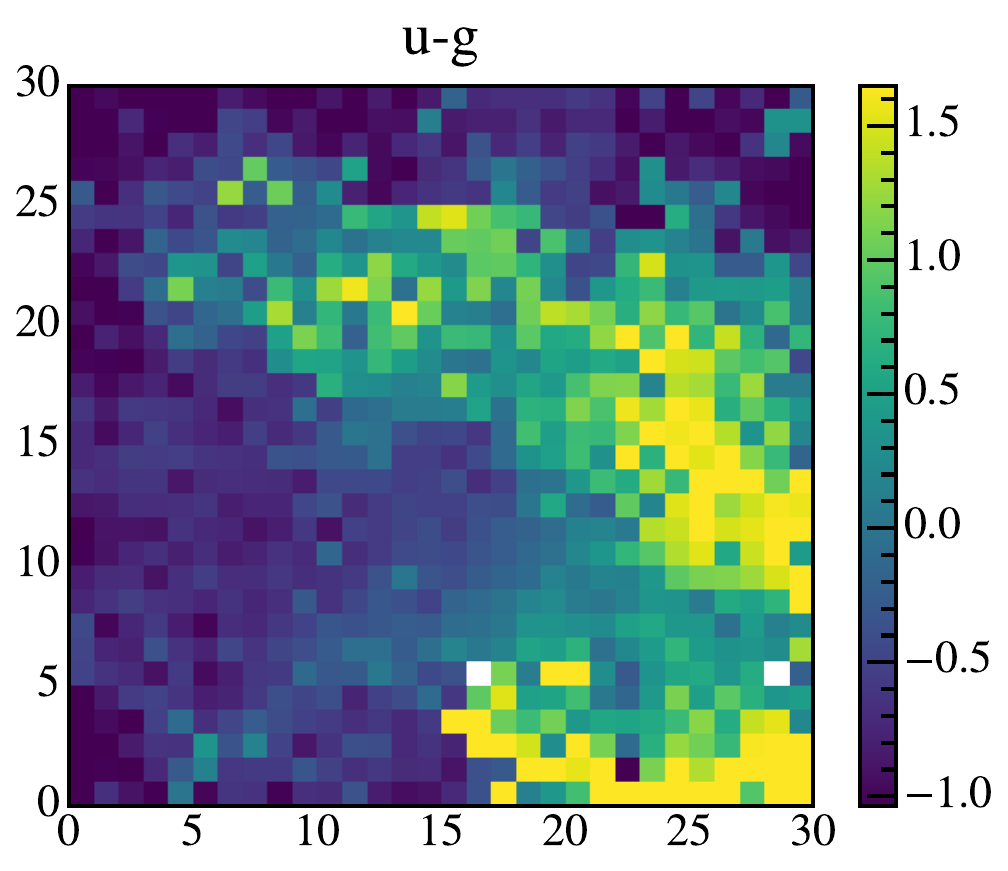}
    \includegraphics[width=0.15\linewidth]{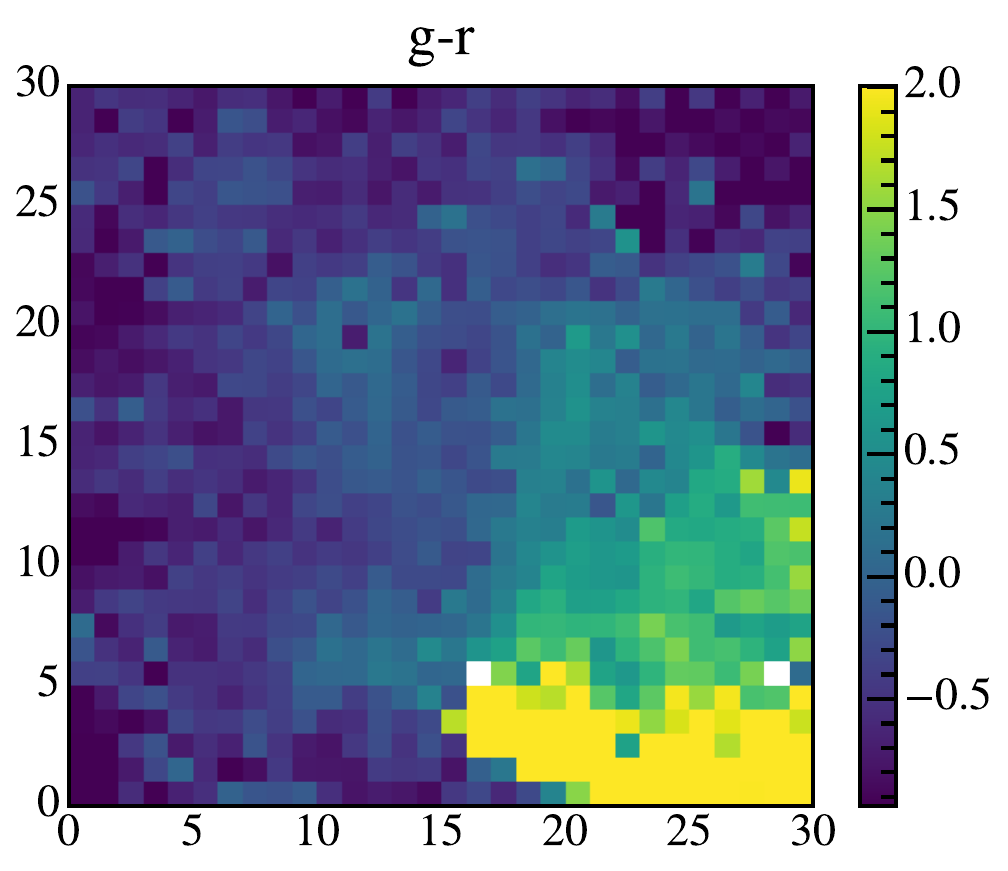}
    \includegraphics[width=0.15\linewidth]{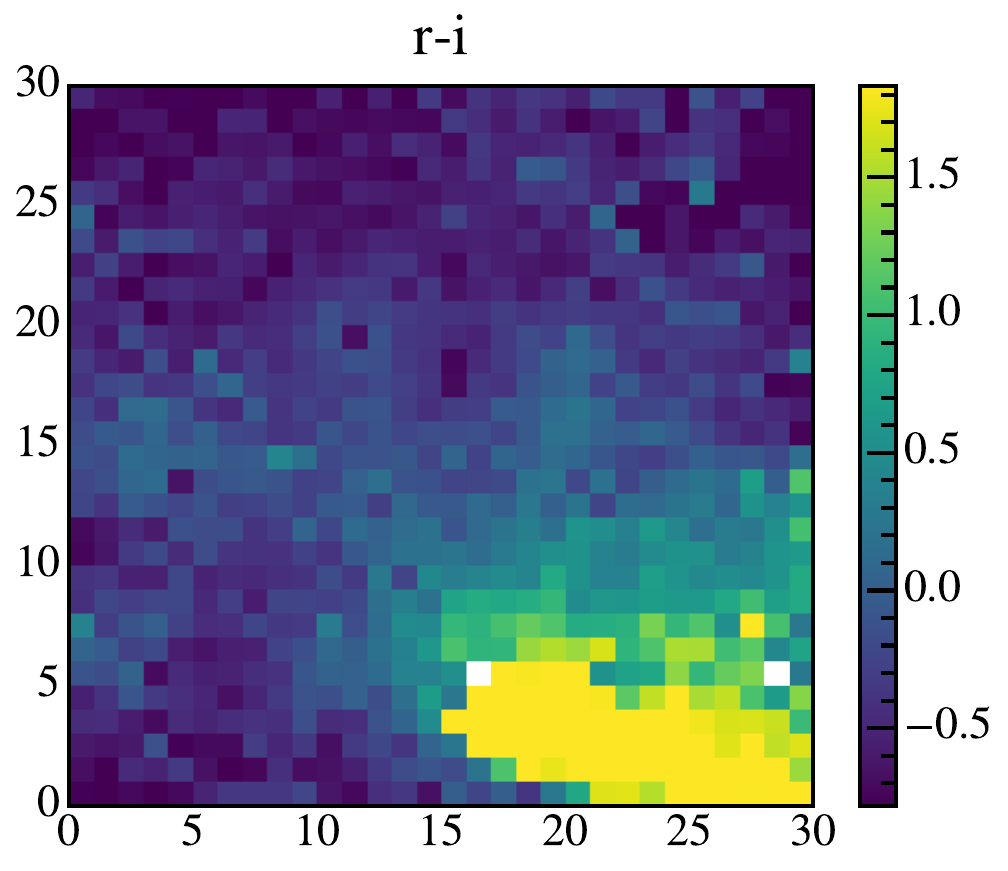}
    \includegraphics[width=0.15\linewidth]{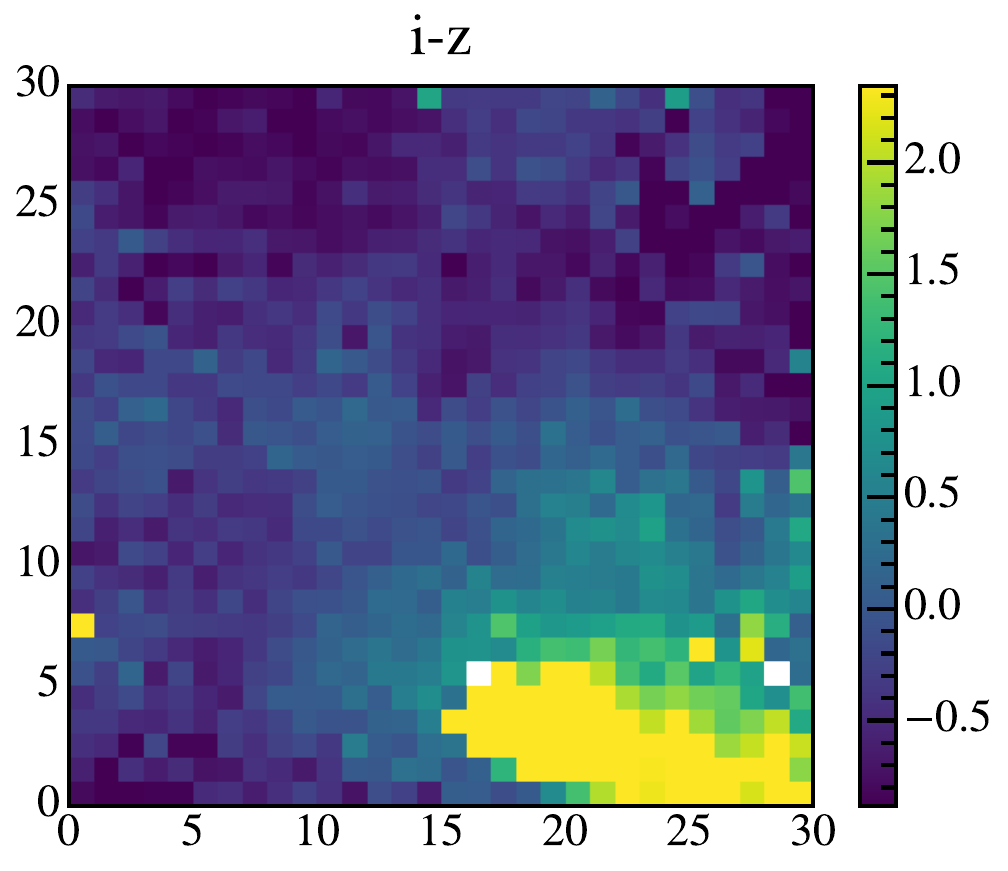}
    \includegraphics[width=0.15\linewidth]{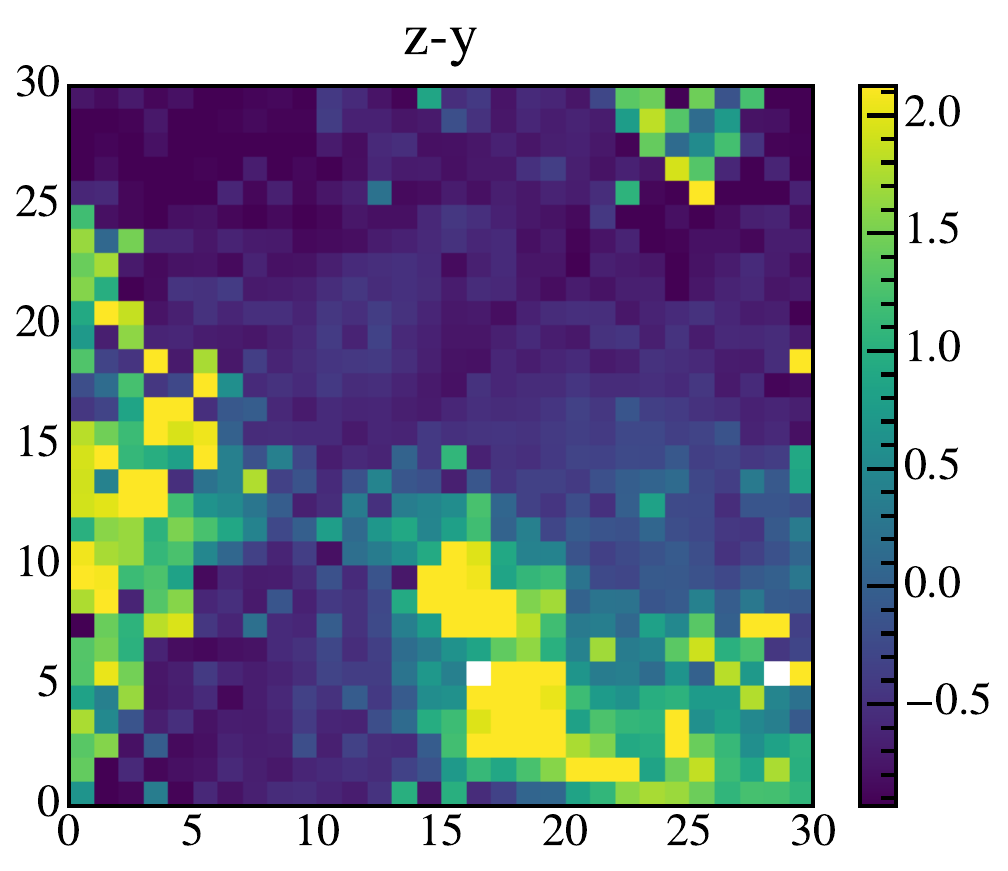}
    \includegraphics[width=0.15\linewidth]{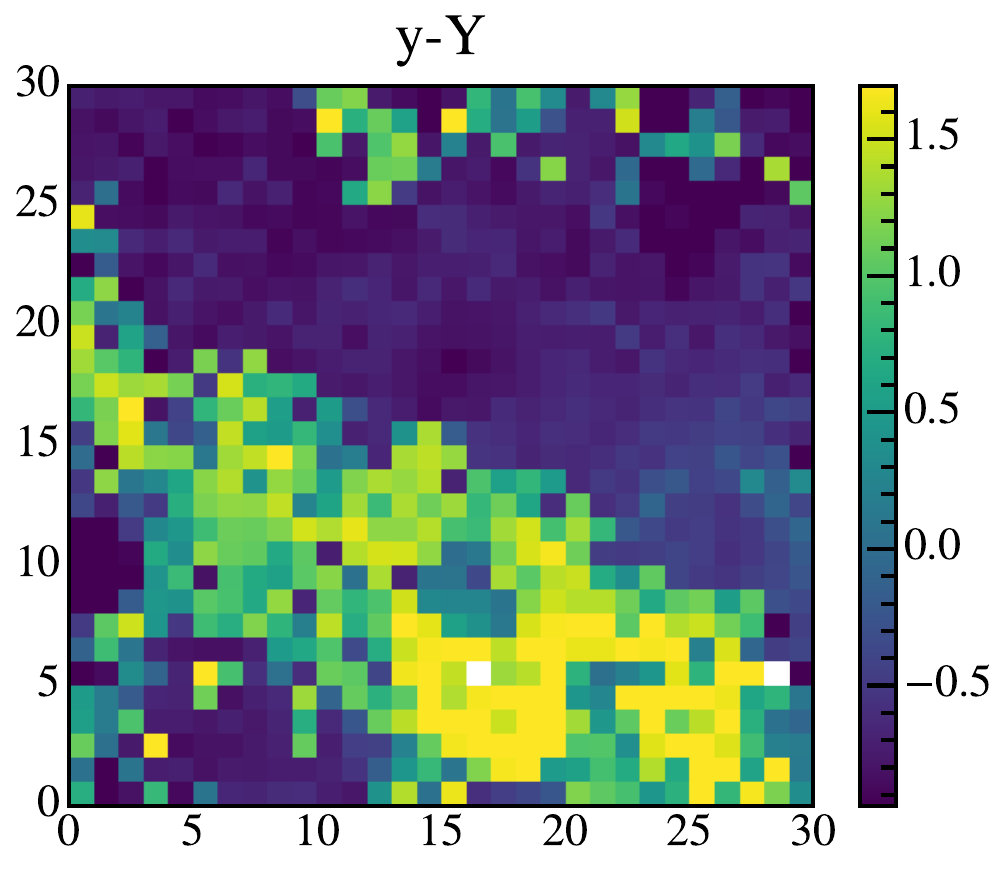}
    \includegraphics[width=0.15\linewidth]{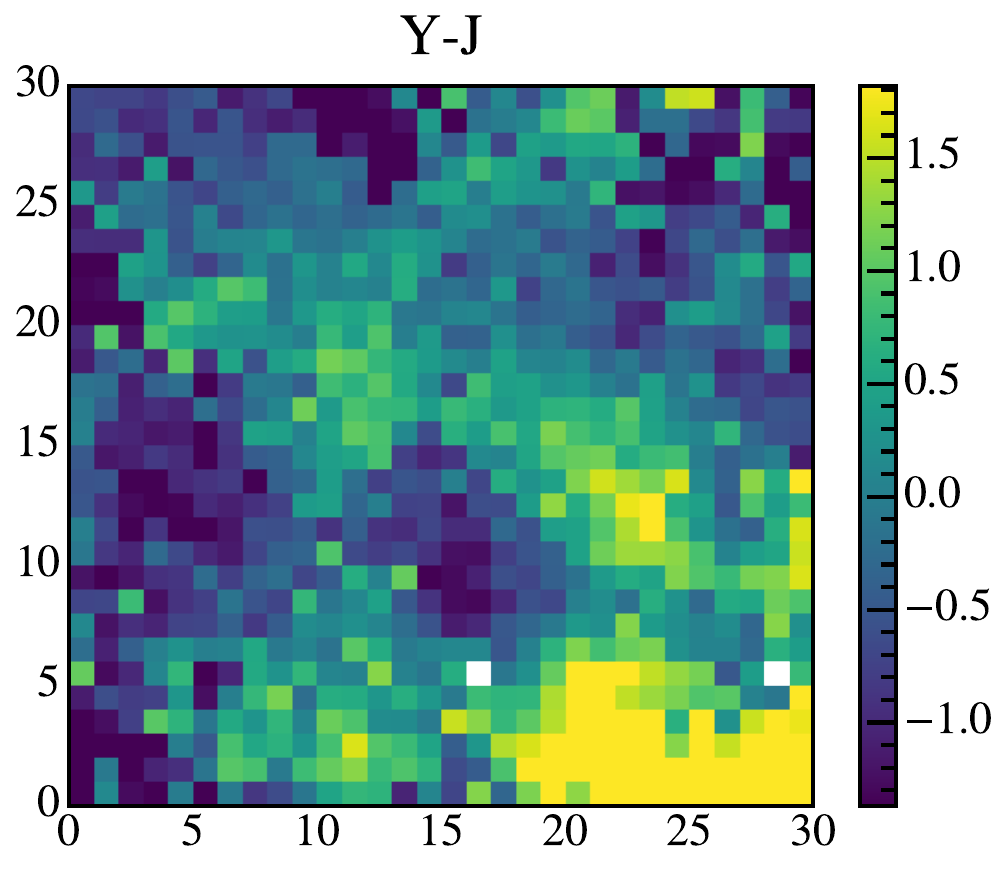}
    \includegraphics[width=0.15\linewidth]{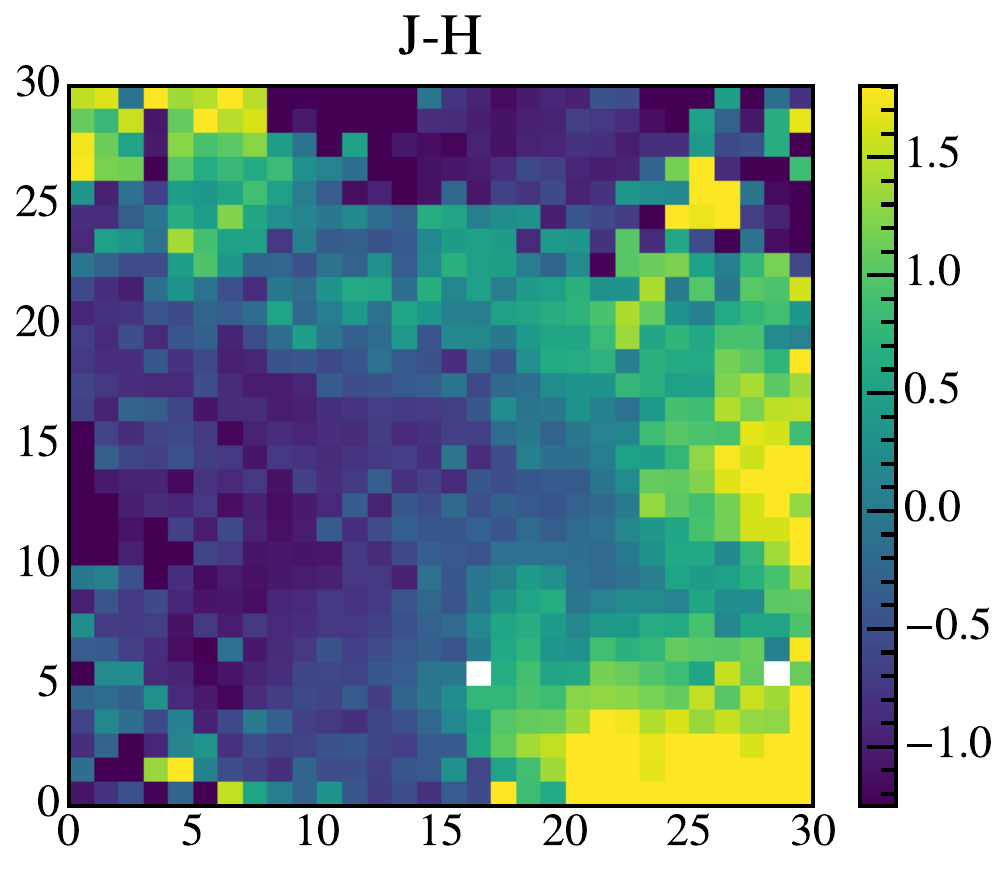}
    \includegraphics[width=0.15\linewidth]{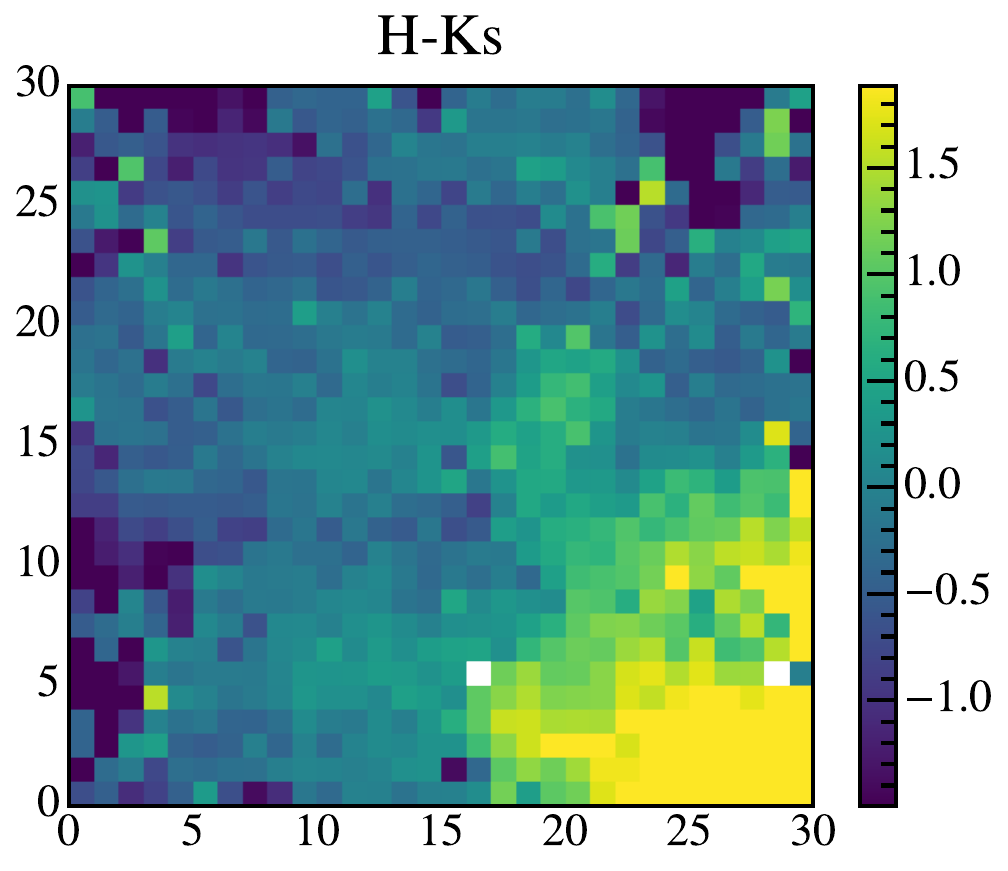}
    \includegraphics[width=0.15\linewidth]{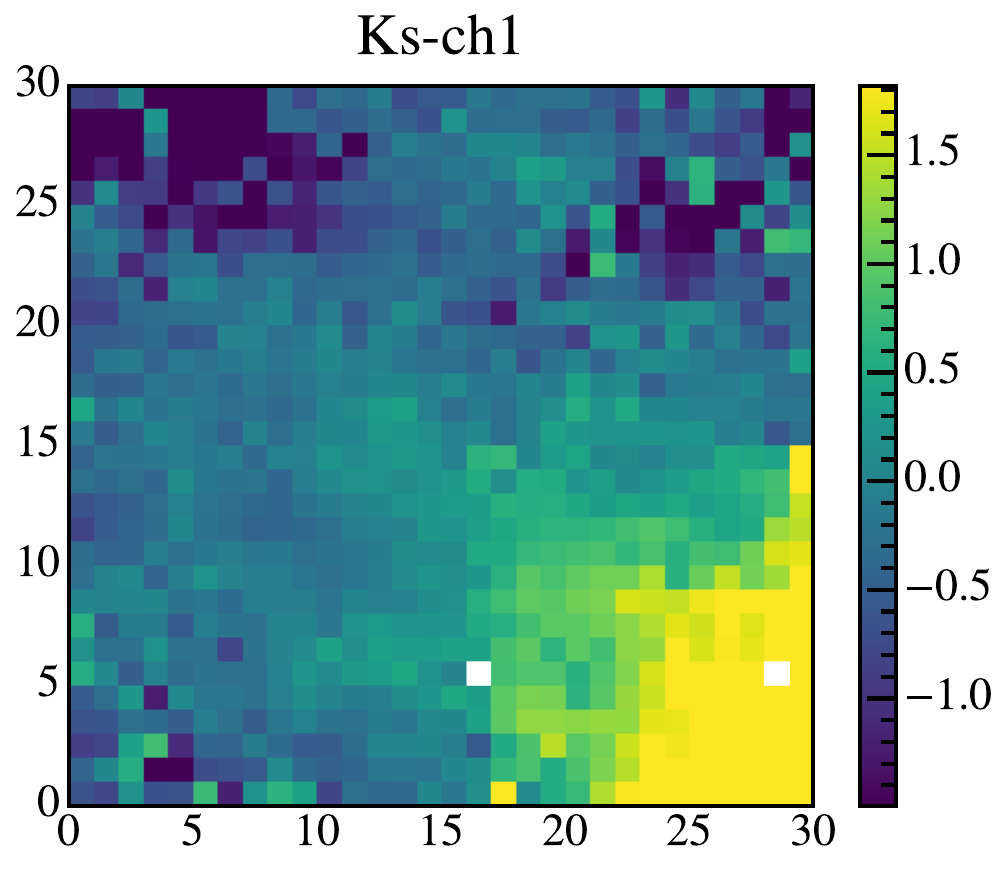}
    \includegraphics[width=0.15\linewidth]{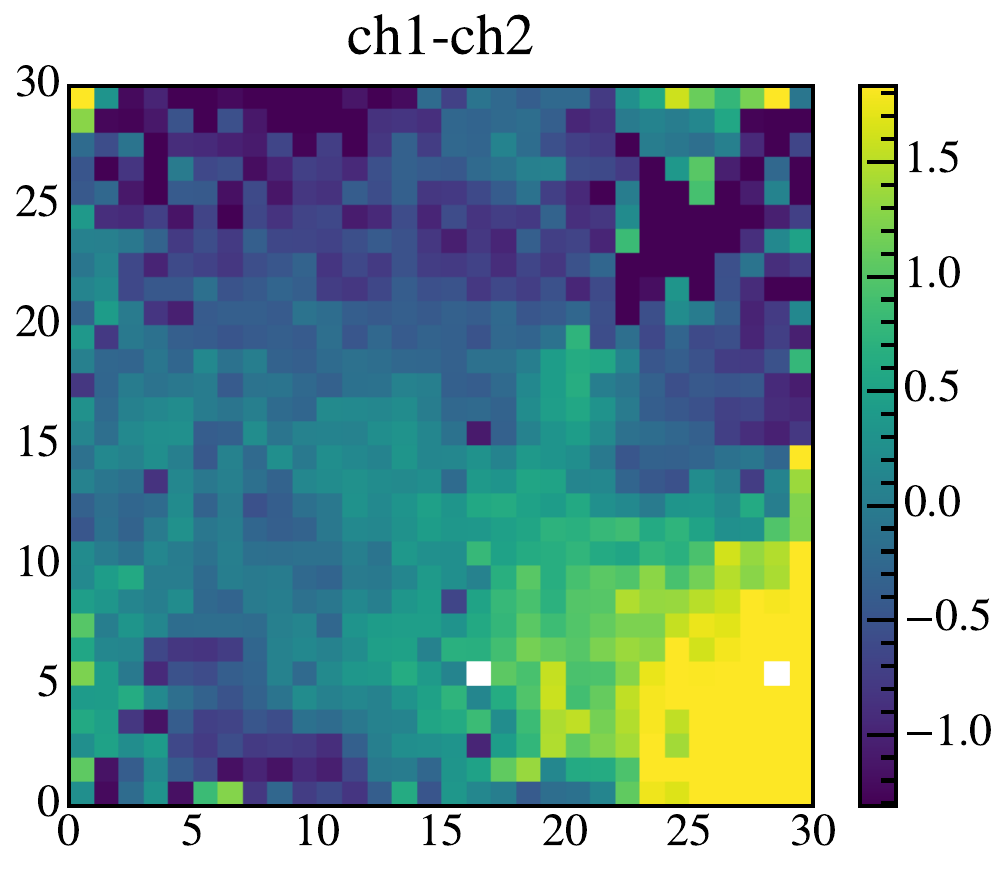}
    
    \caption{Component maps of synthetic $\rm SOM^*$. The color bar indicates the normalized color values.}
    \label{fig:new_som}
\end{figure*}

\end{document}